\documentclass[amsmath,amssymb,aps,10pt,pra,letterpaper,nofootinbib,balancelastpage,notitlepage,twocolumn,floatfix,preprintnumbers]{revtex4-2}
\usepackage{graphicx}	
\usepackage{ragged2e}	
\usepackage{bm}		    
\usepackage{amsmath,graphicx,epsfig}
\usepackage{epstopdf}
\usepackage{euscript}
\usepackage{amsfonts}
\usepackage{amssymb}
\usepackage{float}
\usepackage{xcolor}
\usepackage[caption=false]{subfig}

\usepackage[sort&compress]{natbib}	 	
	\setcitestyle{square,numbers,comma}	
\usepackage[colorlinks=true,urlcolor=blue,linkcolor=black,citecolor=red,bookmarksnumbered=true]{hyperref}	
\newcommand{\orcid}[1]{\href{https://orcid.org/#1}{\,\includegraphics[width=8px]{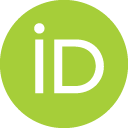}}}

\def\prn#1{{\left(#1\right)}}

\def\sbrk#1{{\left[#1\right]}}
\def\abrk#1{{\langle#1\rangle}}
\def\ket#1{{|#1\rangle}}
\def\bra#1{{\langle#1|}}

\def\dbydt#1{{\frac{d #1}{dt}}}
\def\dbyd#1#2{{\frac{d #1}{d #2}}}
\def\pdbyd#1#2{{\frac{\partial#1}{\partial#2}}}

\def\cg(#1,#2)(#3,#4)(#5,#6){\bra{#1,#2,#3,#4}#5,#6\rangle}

\def\ts#1{{_{\mbox{\scriptsize #1}}}}

\def\threej(#1,#2)(#3,#4)(#5,#6){\begin{pmatrix}#1&#3&#5\\#2&#4&#6\end{pmatrix}}
\def\sixj(#1,#2,#3)(#4,#5,#6){\begin{Bmatrix}#1&#2&#3\\#4&#5&#6\end{Bmatrix}}
\def\ninej(#1,#2,#3)(#4,#5,#6)(#7,#8,#9){\begin{Bmatrix}#1&#2&#3\\#4&#5&#6\\#7&#8&#9\end{Bmatrix}}

\def\sL{{\ensuremath{\EuScript L}}}
\def\sH{{\ensuremath{\EuScript H}}}

\def\bs{\boldsymbol}
\def\mc{\mathcal}

\begin{document}

\title{Quantum dynamics of a levitated ferromagnetic gyroscope}

\date{\today}

\author{Derek F.~Jackson Kimball\orcid{0000-0003-2479-6034}}	
\email{derek.jacksonkimball@csueastbay.edu}
\affiliation{Department of Physics, California State University -- East Bay, Hayward, CA 94542, USA}


\begin{abstract}
We develop a quantum model for the rotational dynamics of a freely floating levitated ferromagnetic gyroscope (LFG), emphasizing the interplay between intrinsic spin $\bs{S}$, mechanical angular momentum $\bs{L}$, and magnetic torque. The conserved total angular momentum projection along the $z$-directed magnetic field $\bs{B}$, $J_z=S_z+L_z$, is quantized, leading in the small-libration-amplitude limit to discrete precessional states $\ket{m}$ (eigenstates of $J_z$ with eigenvalues $J_z = m\hbar$, with $m$ being integer or half-integer) and librational harmonic oscillator states $\ket{n}$ ($n=0,1,2,\ldots$). The discreteness of the energies and dynamical variables is governed by the quantum precession scale $\Omega_Q=\hbar/I$, where $I$ is the moment of inertia of the LFG. We find that the phenomenon of LFG precession persists into high-field regimes where the magnitude of the rotational angular momentum associated with precession exceeds the total intrinsic spin. We analyze the complementary quantum limits of localized semiclassical LFG orientation wave packets and exact $J_z$-eigenstates $\ket{m}$, clarifying the relation between classical precession signals and the underlying quantized spin-rotor dynamics. We further show that radio-frequency fields can drive $\Delta m = \pm 1$ and $\Delta n = \pm 1$ transitions, enabling ladder spectroscopy, tilt-angle control, and sideband-like coupling between precession and libration. The coupled dynamics also exhibit branch-point magnetic resonances where precession and librational motion become strongly coupled. These results establish a framework for using LFGs not only as ultrasensitive torque and magnetic-field sensors, but also as controllable mesoscopic quantum systems. The techniques developed here may be applied to searches for exotic, beyond-the-standard model spin-dependent interactions, ultralight dark matter, and spin-gravity couplings.
\end{abstract}

\maketitle

\section{Introduction}
\label{sec:intro}

In a series of recent studies \cite{kimball2016precessing,fadeev2021ferromagnetic,fadeev2021gravity,vinante2021surpassing,kalia2024ultralight}, a novel regime of quantum spin dynamics in levitated, sub-millimeter-scale hard ferromagnets has been identified: gyroscopic precession driven not by classical rotation, but by the collective intrinsic quantum spin of the magnetized material.
This gyroscopic behavior leads to precession dynamics analogous to those of atomic and nuclear spins in magnetic fields, but now realized in a solid mesoscopic object with strongly correlated electron spins \cite{kimball2016precessing}.
Although this regime was theoretically predicted some time ago \cite{budker2008atomic,kimball2016precessing}, only recently has the first experimental hint of such spin-driven precession been reported in a ferromagnet levitated above a superconductor \cite{ahrens2026observation}.
At present, the gyroscopic spin dynamics of ferromagnets remains an open and largely unexplored frontier.

Further study of the dynamics of such levitated ferromagnetic gyroscopes (LFGs) is motivated by the observation that precision torque measurements using LFGs have the potential to surpass both the standard quantum limit for independent (uncorrelated) spins \cite{kimball2016precessing} and the energy resolution limit \cite{vinante2021surpassing,palacios2022single,ahrens2025levitated} that constrain conventional spin-based quantum sensors \cite{mitchell2020colloquium,jackson2023probing}.
This extraordinary sensitivity arises from the unique physical properties of hard ferromagnets: strong spin-lattice coupling produces a robust, highly ordered spin state in which quantum noise is averaged over a broad frequency bandwidth, dramatically reducing its impact at the low frequencies relevant to precision measurement of gyroscopic spin precession \cite{kimball2016precessing,ni2025microscopic}.
When mechanically isolated from environmental perturbations via levitation, these properties make possible, in principle, torque and magnetic field sensitivities several orders of magnitude beyond existing approaches \cite{jackson2023probing}.
If experimentally realized, this regime would open a new frontier in quantum metrology with wide-ranging implications for fundamental physics \cite{fadeev2021ferromagnetic,fadeev2021gravity,kalia2024ultralight,stickler2021quantum,vinante2022levitated}.

Experiments with ferromagnets levitated above superconductors have demonstrated ultralow mechanical damping and promising torque sensitivity \cite{wang2019dynamics,vinante2020ultralow,ahrens2025levitated,ahrens2026observation}, while related work with levitated micromagnets, spin-mechanical systems, and electrodynamic or magnetic traps has explored the coupling of magnetization to center-of-mass, librational, and rotational degrees of freedom \cite{gieseler2020single,huillery2020spin,perdriat2021spin,perdriat2024rotational,barry2023ferrimagnetic,fuwa2023ferromagnetic,janse2024characterization,ji2025levitated}.
Theoretical studies have clarified the role of gyromagnetic effects, image-dipole back-action for Meissner-effect-based levitation, damping, and microscopic spin-lattice coupling in the classical dynamics of levitated ferromagnets \cite{band2018dynamics,belovs2025gyromagnetic,ni2025microscopic}.

Motivated by these developments, the present work develops an explicitly quantum description of the rotational dynamics of a freely floating LFG.
This provides a framework for LFG magnetic resonance and quantum control, and connects naturally to emerging proposals for macroscopic spin-rotor quantum states in levitated ferromagnets~\cite{ni2026macroscopic,wachter2026gyroscopically}.

\begin{figure}
\center
\includegraphics[width=7.5cm]{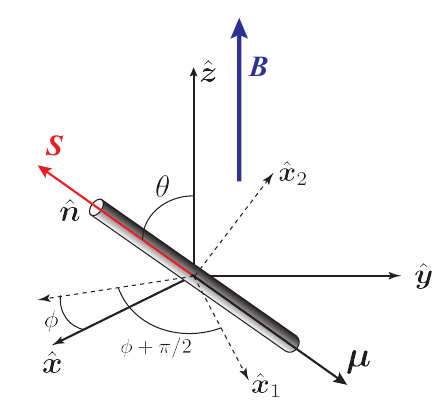}
\caption{Schematic diagram of the geometry used in our model of a levitated ferromagnetic gyroscope (LFG). The LFG is represented by the cylinder. The magnetic field $\bs{B}$ (blue arrow) is along $\hat{\bs{z}}$. The spin $\bs{S}$ (red arrow), pointing along $\hat{\bs{n}}$, is antiparallel to the magnetic moment $\bs{\mu}$. The direction of the unit vector is specified as $\hat{\bs{n}} = \prn{\sin\theta\cos\phi,\sin\theta\sin\phi,\cos\theta}$. $\hat{\bs{x}}_1$ is perpendicular to $\hat{\bs{n}}$ and lies in the $xy$-plane.}
\label{fig:EulerAngles}
\end{figure}

Let us consider an ideal LFG: a freely floating single-domain ferromagnet with macroscopic total intrinsic spin $\bs{S}$.
The macrospin $\bs{S}$ is oriented in a direction $\hat{\bs{n}}$ along the crystalline anisotropy axis of the ferromagnet \cite{chikazumi1997physics} so that $\bs{S}=S\hat{\bs{n}}$.
We assume the net spin takes on its saturation value $S = N\hbar/2$, with $N$ being the number of polarized electrons in the ferromagnet, and also that the angular momentum along $\hat{\bs{n}}$ is dominated by the intrinsic spin $\bs{S}$, neglecting any rotational angular momentum about the axis defined by $\hat{\bs{n}}$.\footnote{If the ferromagnet did have nonzero rotational angular momentum along $\hat{\bs{n}}$, the principal effect would be to decrease the effective gyromagnetic ratio.}
We assume the LFG has cylindrical symmetry and possesses a moment of inertia $I_{1} = I_{2} \equiv I$ and $I_{n} \approx 0$, where the moments of inertia are referenced to the body-centered frame as shown in Fig.~\ref{fig:EulerAngles}, with the $\hat{\bs{x}}_1$ and $\hat{\bs{x}}_2$ directions transverse to $\hat{\bs{n}}$.
This moment of inertia approximates that for a long, thin ferromagnetic needle as originally considered in Ref.~\cite{kimball2016precessing}.
Such a geometry creates a strong shape anisotropy that generates a large optical (zero-wave-vector or uniform precession) magnon gap \cite{frait1965ferromagnetic,diehl2001crystalline}, ensuring maximal spin polarization along $\hat{\bs{n}}$ \cite{seynaeve2001transition} (see further discussion in Sec.\,\ref{sec:numerical-estimates}).

For simplicity, we assume no interaction of the LFG with the external environment except for that with a magnetic field $\bs{B}$ applied along $\hat{\bs{z}}$.
The magnetic field $\bs{B}=B\hat{\bs{z}}$ generates a magnetic torque
\begin{align}
\bs{\tau}_B = \bs{\mu} \times \bs{B} = -g \mu_B \frac{\bs{S}}{\hbar} \times \bs{B}\,,
\label{eq:magnetic-torque}
\end{align}
where $\bs{\mu}$ is the LFG's magnetic moment, $g$ is the Land\'e factor, $\mu_B$ is the Bohr magneton, and the negative sign accounts for the fact that the electron magnetic moment points opposite to its spin.
Application of $\bs{B}$ to the LFG will cause $\bs{\mu} \propto \bs{S}$ to precess and librate (nutate) about $\bs{B}$ \cite{budker2008atomic,kimball2016precessing,fadeev2021ferromagnetic}.
Although generically an LFG must be trapped to confine its center-of-mass motion to a region of space where it can be efficiently interrogated, in the present work we limit our focus to the LFG rotational dynamics and assume the LFG is freely floating.
Future work will explore the effects of a trapping potential.

The central goals of the present work are to (1) elucidate the {\emph{high field}} LFG dynamics, where the rotational angular momentum $L$ of the ferromagnet associated with precession exceeds $S$, (2) explore the {\emph{quantized nature}} of the ferromagnet's precession and libration (nutation), which has implications for understanding and controlling LFG dynamics, and to (3) investigate the possibility of manipulating LFGs using rf fields and magnetic resonance techniques.
From the outset we highlight several notable features of the (quantized) LFG dynamics:
\begin{itemize}

\item{Since there is no component of magnetic torque along $\hat{\bs{z}}$, the projection of the total angular momentum $\bs{J}$ along $\hat{\bs{z}}$, $J_z$, is conserved.}

\item{The conserved angular momentum projection $J_z$ is quantized and takes on discrete values $J_z = m\hbar$ where $m$ is integral (or half-integral).}

\item{Due to the magnetic torque $\bs{\mu} \times \bs{B}$ and conservation of $J_z$, the rotational angular momentum $\bs{L}$ associated with LFG precession is coupled to the spin $\bs{S}$, and the coupled spin and rotational quantum states fulfill the condition $J_z = S_z + L_z = m\hbar$.}

\item{As noted in Ref.~\cite{kimball2016precessing}, spin fluctuations transverse to the $\hat{\bs{n}}$ direction are rapidly averaged compared to the macroscopic dynamics of the LFG, and therefore $\bs{S}$ can be treated as a \emph{classical} variable locked to the lattice direction $\hat{\bs{n}}$.}

\item{LFG precession is {\emph{not}} restricted to the low-field regime identified in the original precessing-needle-magnetometer proposal of Ref.~\cite{kimball2016precessing}, namely where the Larmor frequency is below a threshold $\Omega \ll \Omega^\star \equiv \omega_I$, with $\omega_I = S/I$ being the Einstein-de Haas frequency. We find that precessional dynamics persist into regimes where the rotational angular momentum associated with precession exceeds the intrinsic spin, $L \gtrsim S$.}

\item{Near the equator ($\theta = \pi/2$), the quantized LFG dynamics can be described by coupled ladders of $J_z$ eigenstates $\ket{m}$ and librational harmonic oscillator states $\ket{n}$, where $n = 0,1,2,\ldots$. Similar quantum dynamics for small amplitude libration occur at all tilt angles $\theta$.}


\item{There are two complementary quantum regimes for LFG precession due to an uncertainty relation between $J_z$ and the azimuthal angle $\phi$. A semiclassical precession signal corresponds to a state localized in $\phi$, and hence to a superposition of many $J_z$ eigenstates. Conversely, an exact eigenstate of $\hat{J}_z$ corresponds to a delocalized azimuthal angle $\phi$ and does not by itself produce a classical transverse precession signal.}

\item{Radio-frequency (rf) fields can drive transitions between these quantized levels: circularly polarized rf fields couple neighboring $m$-levels through $\Delta m = \pm 1$ transitions, while linearly polarized fields can drive neighboring librational levels through $\Delta n = \pm 1$ transitions. Thus rf fields provide a route to coherent control and spectroscopy of LFG quantum dynamics.}

\item{The coupled precession-libration dynamics exhibit branch-point magnetic resonances, where the ``fast'' and ``slow'' no-nutation precession branches merge. At such points the precession frequency becomes extremely sensitive to changes in the tilt angle $\theta$, leading to enhanced coupling between precession and libration.}

\item{The existence of both $m$- and $n$-ladders allows sideband-like resonances between precessional and librational degrees of freedom. In the presence of suitable rf perturbations, these resonances may enable exchange of energy between precession and libration, analogous to sideband transitions in other quantum systems.}

\item{The discreteness of the LFG quantum dynamics is governed by the quantum precession frequency scale $\Omega_Q=\hbar/I$, which increases rapidly as the LFG size is reduced.}

\item{Cooling the orientational degrees of freedom to low occupation appears feasible in principle by using high-field polar libration to create a larger energy gap, applying sideband cooling, and then adiabatically ramping the magnetic field while maintaining isolation from environmental thermalization.}

\end{itemize}

The remainder of this paper is organized as follows.
In Sec.\,\ref{sec:classical-model} we develop a classical model of the LFG dynamics, which turn out to be analogous to the well-known dynamics of a heavy symmetric top.
In Sec.\,\ref{sec:quantum-model} we introduce the corresponding quantum Hamiltonian and discuss the treatment of the ferromagnetic macrospin $\bs{S}$ as a classical vector locked to the crystal lattice.
In Sec.\,\ref{sec:precession-small-libration} we analyze the quantized precession and libration dynamics in several important limits, including equatorial precession, polar libration, and rf-driven ladder spectroscopy.
In Sec.\,\ref{sec:magnetic-resonance} we discuss magnetic-resonance phenomena associated with LFG dynamics, including rf control of $J_z$, branch-point resonances, pole avoidance, and sideband-like couplings between precessional and librational levels.
In Sec.\,\ref{sec:numerical-estimates} we present numerical estimates for experimentally relevant LFG parameters and discuss the prospects for observing quantized dynamics.
In Sec.\,\ref{sec:cooling} we outline possible cooling strategies, focusing on high-field polar preparation followed by isolation and adiabatic field ramping.
We conclude in Sec.\,\ref{sec:conclusion} with a summary and outlook for future experimental and theoretical work.

\section{Classical model of a levitated ferromagnetic gyroscope}
\label{sec:classical-model}

We describe the dynamics of the LFG using the Euler angles $\phi$ and $\theta$, which are equivalent to the usual spherical coordinate angles $\phi$ and $\theta$ that describe the direction of the rotated LFG axis, as shown in Fig.~\ref{fig:EulerAngles}.
The precessional motion is described by $\phi(t)$ and the librational motion (also known as nutation) is described by $\theta(t)$.\footnote{The term {\emph{nutation}} refers to the gyroscope-like oscillation of the tilt angle superposed on precession, and {\emph{libration}} refers to the small-angle oscillations about a stable equilibrium orientation of the LFG; in the present work, following Refs.~\cite{fadeev2021ferromagnetic,vinante2021surpassing,fadeev2021gravity,kalia2024ultralight}, these terms are used interchangeably in the description of the LFG dynamics.}
The system is characterized by the Lagrangian
\begin{align}
\sL = \frac{I}{2} \prn{ \dot\theta^2 + \sin^2\theta \dot\phi^2 } + S\cos\theta \dot\phi - S\Omega \cos\theta\,,
\label{eq:Lagrangian}
\end{align}
where
\begin{align}
\Omega = \frac{ g \mu_B B }{\hbar}
\end{align}
is the Larmor frequency.
The term
\begin{align}
K = \frac{I}{2} \prn{ \dot\theta^2 + \sin^2\theta \dot\phi^2 }
\label{eq:kinetic-term}
\end{align}
describes the rotational kinetic energy of the LFG, the potential energy of the ferromagnet in the magnetic field $\bs{B}$ is
\begin{align}
U = -\bs{\mu} \cdot \bs{B} = \frac{g \mu_B}{\hbar} S B \cos\theta = S\Omega \cos\theta\,,
\label{eq:potential-energy}
\end{align}
and
\begin{align}
\sL\ts{WZ} = S\cos\theta \dot\phi
\label{eq:Wess-Zumino-term}
\end{align}
is the Wess-Zumino term \cite{loss1992suppression} associated with the geometric (Berry's) phase acquired by the spins during precession \cite{berry1984quantal,arovas1988functional}, see Appendix~\ref{app:Wess-Zumino-term}.
Crucially, as we will see, for the LFG it is the Wess-Zumino term that couples spin to precession and enforces the conservation of $J_z$.

From Eq.~\eqref{eq:Lagrangian}, we find the canonical momenta
\begin{align}
p_\theta = \pdbyd{\sL}{\dot{\theta}} = I\dot\theta = J_1
\label{eq:p-theta}
\end{align}
and
\begin{align}
p_\phi = \pdbyd{\sL}{\dot{\phi}} = I\sin^2\theta\dot\phi + S\cos\theta = J_z\,,
\label{eq:p-phi}
\end{align}
where we can identify $p_\theta = J_1$ as the angular momentum projection along the $x_1$ axis (Fig.~\ref{fig:EulerAngles}) and $p_\phi = J_z$ as the angular momentum projection along $z$, which accounts for both the spin $S_z$ and the rotational angular momentum $L_z$ associated with precession.
Equation~\eqref{eq:p-phi} makes evident the role played by the Wess-Zumino term [Eq.~\eqref{eq:Wess-Zumino-term}] in describing the spin-orbit coupling.

The Hamiltonian for the LFG is obtained from
\begin{align}
\sH = p_\theta \dot\theta + p_\phi \dot\phi - \sL
\label{eq:Hamiltonian-general-formula}
\end{align}
and is found to be, after some algebra,
\begin{align}
\sH = \frac{1}{2I}p_\theta^2 + \frac{1}{2 I \sin^2\theta} \prn{ p_\phi - S\cos\theta }^2 + S\Omega\cos\theta\,.
\label{eq:Hamiltonian}
\end{align}

We can obtain the equations of motion for the LFG from Hamilton's equations:
\begin{align}
\dbydt{p_\theta} & = -\pdbyd{\sH}{\theta}\,, \label{eq:Hamilton-Eqn-1} \\
\dbydt{p_\phi} & = -\pdbyd{\sH}{\phi}\,. \label{eq:Hamilton-Eqn-2}
\end{align}
Based on the above equations, after some algebra, we find that
\begin{align}
\ddot{\theta} - \dot{\phi}^2\sin\theta\cos\theta + \omega_I\dot{\phi}\sin\theta & = \Omega \omega_I \sin\theta~, \label{eq:Eqn-of-motion-1} \\
\ddot{\phi}\sin\theta + 2\dot{\phi}\dot{\theta}\cos\theta - \omega_I\dot{\theta} & = 0~. \label{eq:Eqn-of-motion-2}
\end{align}
where we have identified
\begin{align}
\omega_I = \frac{S}{I}
\label{eq:Einstein-de-Haas-frequency}
\end{align}
as the Einstein-de Haas frequency \cite{Ein1915spinrotation1,Ein1915spinrotation2}.
The condition for precession-dominated dynamics identified in Ref.\,\cite{kimball2016precessing} is
\begin{align}
\Omega \ll \omega_I\,,
\label{eq:precession-dominated-regime-condition}
\end{align}
although, as we will see in Sec.\,\ref{sec:precession-small-libration}, precession-dominated dynamics are not only limited to the regime identified by Eq.\eqref{eq:precession-dominated-regime-condition}, but can also be observed for $\Omega > \omega_I$.
This opens more applications of LFGs as sensors to measure larger magnetic fields and torques than envisioned, for example, in Refs.\,\cite{kimball2016precessing,fadeev2021ferromagnetic}.
We note that Eqs.~\eqref{eq:Eqn-of-motion-1} and \eqref{eq:Eqn-of-motion-2} are analogous to the equations of motion for the textbook example of a heavy symmetric top (see, for example, Ref.~\cite{morin2008introduction}).
Related classical gyromagnetic dynamics of levitated magnetic particles, including Einstein-de Haas and Barnett effects, were recently derived in Ref.~\cite{belovs2025gyromagnetic}, where the dissipationless ferromagnetic case was also formulated in Hamiltonian form and used to analyze precession and nutation.

\section{Quantum model of a levitated ferromagnetic gyroscope}
\label{sec:quantum-model}

A quantum model of the LFG is obtained by describing the angular position of $\bs{S}$ using angle operators $\hat{\theta}$ and $\hat{\phi}$ and introducing the momentum operators
\begin{align}
\hat{p}_\theta &= -i\hbar \pdbyd{}{\theta} = \hat{J}_1\,, \label{eq:p-theta-op} \\
\hat{p}_\phi &= -i\hbar \pdbyd{}{\phi} = \hat{J}_z \,. \label{eq:p-phi-op}
\end{align}
The Hamiltonian operator for the LFG system is thus
\begin{align}
\hat{H} = \frac{1}{2I}\hat{p}_\theta^2 + \frac{1}{2 I \sin^2\hat{\theta}} \prn{ \hat{p}_\phi - S\cos\hat{\theta} }^2 + S\Omega\cos\hat{\theta}\,.
\label{eq:Hamiltonian-op}
\end{align}
Note that
\begin{align}
\sbrk{\hat{p}_\phi,\hat{H}} = \sbrk{\hat{J}_z,\hat{H}}  =0\,,
\label{eq:p-phi-H-commutator}
\end{align}
since $\hat{\phi}$ does not appear in $\hat{H}$.
Therefore, as noted earlier, $\hat{p}_\phi=\hat{J}_z$ is a conserved quantity.
Furthermore, because of Eqs.~\eqref{eq:p-theta-op} and \eqref{eq:p-phi-op},
\begin{align}
\sbrk{\hat{\theta},\hat{p}_\theta} = \sbrk{\hat{\theta},\hat{J}_1} & = i\hbar \,, \label{eq:theta-p-theta-commutator} \\
\sbrk{\hat{\phi},\hat{p}_\phi} = \sbrk{\hat{\phi},\hat{J}_z} & = i\hbar \,, \label{eq:phi-p-phi-commutator}
\end{align}
meaning that the angular positions and momenta are complementary observables that obey Heisenberg uncertainty relations.

The quantum version of Eq.~\eqref{eq:p-phi} yields an expression for the operator describing the precession frequency,
\begin{align}
\hat{\omega} = \hat{\dot{\phi}} = \frac{\hat{J}_z-S\cos\hat{\theta}}{I\sin^2\hat{\theta}} = \frac{\hat{L}_z}{I\sin^2\hat{\theta}}\,.
\label{eq:precession-op}
\end{align}

A crucial point in the above quantization of the LFG system is that we have described the macrospin $\bs{S}$ \cite{xiao2005macrospin} of the LFG as a \emph{classical} quantity locked to the crystalline anisotropy axis pointing along $\hat{\bs{n}}$,
\begin{align}
\bs{S} \equiv \abrk{\bs{S}} = S\hat{\bs{n}}\,,
\label{eq:classical-spin}
\end{align}
instead of as an operator.
This description goes to the heart of the argument of Ref.~\cite{kimball2016precessing} explaining how an LFG can, in principle, far surpass the spin-projection noise limit for independent particles.
In the context of the quantum model of an LFG described in the present work, it is worth reconsidering this key point.
The dynamics we are interested in are driven by the action of the magnetic torque \eqref{eq:magnetic-torque} upon $\bs{S}$, and $\bs{S}$ interacts with the lattice to keep $\hat{\bs{n}}$ locked along $\bs{S}$.
If $\bs{S}$ is temporarily tilted away from $\hat{\bs{n}}$, the spin-lattice interaction exerts a restoring torque.
These dynamics correspond to magnon excitations at the ferromagnetic resonance frequency, which in the considered system (a ferromagnet with strong shape anisotropy) is in the GHz--THz range \cite{frait1965ferromagnetic,diehl2001crystalline}.
Furthermore, Landau-Lifshitz-Gilbert damping \cite{landau1935theory,gilbert2004phenomenological} of such excitations occurs at rates far exceeding a MHz \cite{kambersky2007spin}.
Indeed, recent investigations of the closely related Einstein-de Haas and Barnett effects demonstrate that angular momentum exchange between spins and the crystalline lattice of a ferromagnet occurs on the femtosecond timescale \cite{mentink2019quantum}.
Thus the internal microscopic spin fluctuation dynamics of the ferromagnet occur at vastly different frequencies as compared to the frequencies associated with the macroscopic dynamics of the LFG we investigate in the present work \cite{band2018dynamics}.
While there is rapid exchange of angular momentum between the spins and the lattice, Landau-Lifshitz-Gilbert damping quickly relaxes the system to the ground energy state while preserving the total angular momentum, thereby maintaining the macrospin $\bs{S}$ orientation along $\hat{\bs{n}}$.
This is, indeed, the essence of being a ferromagnet: the macroscopic spin polarization is, essentially, preserved for time scales long compared to any observation (for further discussion, including numerical estimates, see Sec.\,\ref{sec:numerical-estimates}).

In the measurement schemes carried out in practice, see, e.g., Refs.~\cite{wang2019dynamics,vinante2020ultralow,gieseler2020single,huillery2020spin,ahrens2025levitated,ahrens2026observation}, any such high-frequency spin fluctuation dynamics, described by the fluctuation-dissipation theorem \cite{brown1963thermal}, are averaged out.
The microscopic physics of the spin-lattice interaction in an LFG has recently been studied in detail in Ref.~\cite{ni2025microscopic}, confirming the above description and justifying the use of the expectation value for the macrospin in our quantum model, Eq.~\eqref{eq:classical-spin}.

\section{Precession in the small libration amplitude limit}
\label{sec:precession-small-libration}

Insight into the dynamics of the LFG can be obtained in the limit where the system is in an eigenstate of the librational motion.\footnote{Note that this is a special case; generic initial conditions will lead to an LFG undergoing combined precession and libration (nutation).}
This means that the LFG axis has a fixed expectation value of the angle $\theta$ with respect to the $z$-axis throughout its dynamics, and we set $\abrk{\dot{\theta}} = 0$ and $\abrk{\ddot{\theta}} = 0$.
In this case we find from Eq.~\eqref{eq:Eqn-of-motion-2} that
\begin{align}
\abrk{\ddot{\phi}} = 0~,
\end{align}
and consequently that $\abrk{\dot{\phi}} = \omega$ where $\omega$ is a constant precession frequency.
(These conditions correspond to the classical criteria for stable precession of a gyroscope with no nutation \cite{morin2008introduction}.)
From Eq.~\eqref{eq:Eqn-of-motion-1}, assuming that $\sin\theta \neq 0$ (i.e., away from the poles at $\theta = 0$ and $\theta=\pi$), we find:
\begin{align}
\cos\theta \omega^2 - \omega_I \omega + \Omega \omega_I = 0\,,
\label{eq:quadratic-eq-for-precession-freq}
\end{align}
which is a quadratic equation for $\omega$.
Solving Eq.~\eqref{eq:quadratic-eq-for-precession-freq}, we obtain two solutions:
\begin{align}
\omega = \frac{\omega_I}{2\cos\theta} \prn{ 1 \pm \sqrt{1-\frac{4\Omega\cos\theta}{\omega_I}} }\,.
\label{eq:precession-frequency-constant-theta}
\end{align}

In the limit where $\Omega \ll \omega_I$, which is the condition \eqref{eq:precession-dominated-regime-condition} for precession-dominated LFG dynamics originally identified in Refs.~\cite{budker2008atomic,kimball2016precessing},
\begin{align}
\omega \approx \frac{\omega_I}{\cos\theta} - \Omega,\,\Omega\,.
\label{eq:precession-frequency-low-field-limit}
\end{align}
The $\omega = \Omega$ case corresponds to the original concept of a precessing LFG described in Ref.~\cite{kimball2016precessing} as discussed in Sec.~\ref{subsec:equator-precession}, corresponding to ``slow'' precession.
Note that if $\bs{S}$ is tilted by $\theta = \pi/2$ relative to $\bs{B}$ and precesses at the equator, $\cos\theta \rightarrow 0$, and so we find that $\omega = \Omega$ for all values of $B$ since the first term in Eq.~\eqref{eq:quadratic-eq-for-precession-freq} $\rightarrow 0$.
The $\omega = \omega_I/\cos\theta - \Omega$ case corresponds to ``fast'' precession of $\bs{S}$ around $\bs{J}$ when the LFG has acquired nonzero $\bs{L}$ even at $\bs{B}=0$.\footnote{Note that there is a special condition on the relationship between the tilt angle $\theta$ and the magnitude of $\bs{L}$ in order to have zero nutation. As a concrete example consider the $\theta=3\pi/4$ case with no external magnetic torque, $\Omega=0$, and with $J_z=0$. Then, according to Eq.\,\eqref{eq:p-phi}, $\omega = \sqrt{2} \omega_I$ is a solution, which indicates a specific $L_z$ value is needed. The $\omega = \Omega$ case corresponds to an LFG ``starting from rest.'' More generally, the LFG dynamics are expected to exhibit both precession and nutation (libration) as the LFG will likely start with a nonzero, random total angular momentum $\bs{J}$.}

\begin{figure}
\center
\includegraphics[width=7.5cm]{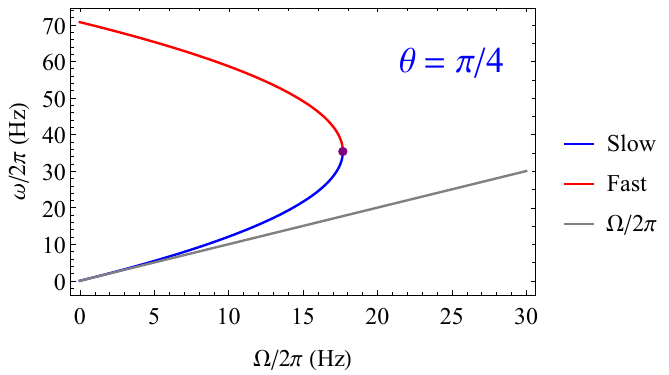}\\[0.3cm]
\includegraphics[width=7.5cm]{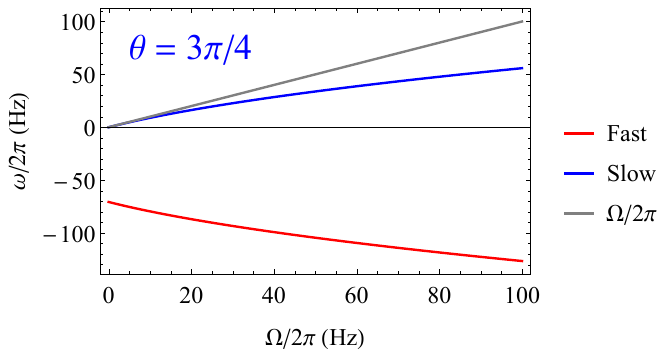}
\caption{The precession frequency $\omega/(2\pi)$ as a function of the Larmor frequency $\Omega/(2\pi)$ for a fixed angle of the LFG axis with respect to the vertical $z$-axis (along $\bs{B}$) for the case of no libration (i.e., no nutation). The Einstein-de Haas frequency is chosen to be $\omega_I = 2\pi \times 50$\,Hz. The upper plot shows the behavior for positive $\cos\theta$, specifically for $\theta=\pi/4$, where the fast precession (the red curve starting at $\omega = \omega_I/\cos\theta$ for $\Omega=0$) and the slow precession (the blue curve starting at $\omega=0$ for $\Omega = 0$) converge at the ``branch point'' where $\Omega=\omega_I/(4\cos\theta)$ and $\omega=\omega_I/(2\cos\theta)$ (marked by the purple dot). The lower plot shows the behavior for negative $\cos\theta$, specifically for $\theta=3\pi/4$ (the fast precession, shown by the red curve, is counter-rotating to the Larmor precession, and described by a negative $\omega$ in our model). The gray lines show $\omega = \Omega$, which is the equatorial ($\theta=\pi/2$) LFG precession frequency for all values of $B$. Note the different scales for the horizontal axes in the upper and lower plots.}
\label{fig:precession-vs-Larmor-freq}
\end{figure}

Another case of interest is the high-field limit of Eq.~\eqref{eq:precession-frequency-constant-theta}, not originally appreciated in Ref.~\cite{kimball2016precessing}, where $\Omega \gg \omega_I$.
If $\cos\theta$ is positive, there appears to be a ``branch point'' in the precession frequency when $\Omega \rightarrow \omega_I/(4\cos\theta)$,
\begin{align}
\omega \rightarrow \frac{\omega_I}{2\cos\theta}\,,
\label{eq:precession-frequency-high-field-asymptote}
\end{align}
corresponding to the situation where the argument inside the square root appearing in Eq.\,\eqref{eq:precession-frequency-constant-theta} goes to zero.
This is the point at which the ``fast'' and ``slow'' precession branches merge as seen in the upper plot of Fig.\,\ref{fig:precession-vs-Larmor-freq}.\footnote{This branch point has a square-root structure reminiscent of exceptional points in non-Hermitian systems. However, in the no-nutation model considered here it is more properly understood as a ``fold bifurcation'' of the steady-precession solution. Damping and relaxation in practical realizations of LFG systems may produce exceptional-point-like behavior, but establishing this requires further analysis.}
At this tilt angle $\theta$, for magnetic fields at or beyond the critical field where $\Omega \geq \omega_I/\prn{4\cos\theta}$, the magnetic torque is now too large to be balanced by the gyroscopic terms in a way that keeps $\theta$ constant.
In this case, the assumptions that $\abrk{\dot{\theta}} = 0$ and $\abrk{\ddot{\theta}} = 0$ cannot be satisfied and so $\theta$ must change and libration/nutation necessarily develops.
This leads to a magnetic resonance further discussed in Sec.\,\ref{sec:magnetic-resonance}.

On the other hand, if $\cos\theta$ is negative, then for $\Omega \gg \omega_I$
\begin{align}
\omega \approx \pm \sqrt{ \omega_I \Omega }\,.
\label{eq:precession-frequency-high-field-limit}
\end{align}
In the equatorial case of $\theta=\pi/2$ where $\cos\theta = 0$, there is a single unique solution of Eq.\,\eqref{eq:quadratic-eq-for-precession-freq} where $\omega=\Omega$ for all values of $\Omega$, as noted above.
These cases are illustrated in Fig.\,\ref{fig:precession-vs-Larmor-freq}, where the dependence of $\omega$ on the Larmor frequency $\Omega$ is plotted for constant $\theta$.
This demonstrates that LFG precessional dynamics can, within the rigid-macrospin model developed here, be present at \emph{magnetic field values much larger than the threshold field} identified in Ref.~\cite{kimball2016precessing}.\footnote{This analysis applies up to values of $\Omega$ where the frequencies of internal magnetic modes are approached, where the rigid-macrospin model breaks down. This occurs at the $\sim$\,GHz scale for many ferromagnetic materials.}

\begin{figure}
\center
\includegraphics[width=7.5cm]{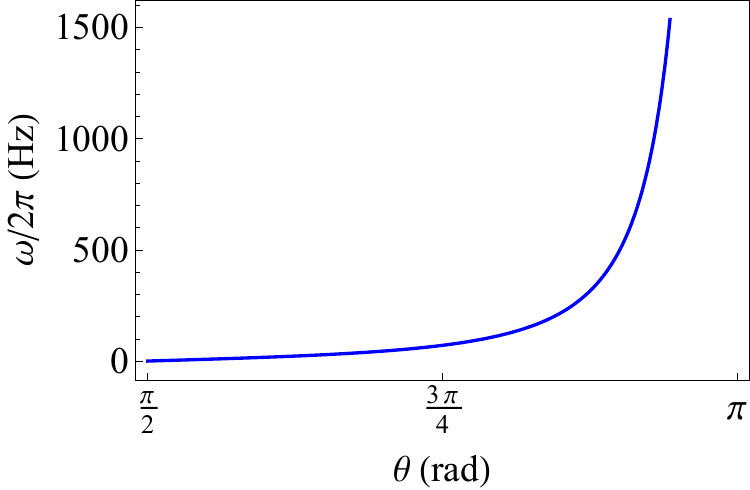}
\caption{The LFG precession frequency $\omega$ as a function of the tilt angle $\theta$ for the case of $J_z=0$.}
\label{fig:omega-vs-theta}
\end{figure}

As noted earlier, the $z$-projection of the total angular momentum $J_z$ is conserved, even for changing $\Omega$.
Due to the conservation of $J_z$, it turns out that for an LFG starting from some particular angle $\theta$, as $B$ increases (thereby increasing $\Omega$), the angle $\theta$ will, necessarily, change, so that the change of $L_z$ is compensated by a corresponding change of $S_z = S\cos\theta$.
Practically, the case of constant $J_z$ is the more physically relevant scenario as compared to the case of constant $\theta$, since $J_z$ is a conserved quantity.
For simplicity, consider the case for which $J_z=0$ and therefore $L_z = -S_z$, in which case, based on Eq.\,\eqref{eq:precession-op},
\begin{align}
\omega = -\omega_I \frac{\cos\theta}{\sin^2\theta}\,,
\label{eq:prec-freq-vs-angle-Jz-is-zero}
\end{align}
demonstrating the relationship between precession frequency and $\theta$, plotted in Fig.~\ref{fig:omega-vs-theta}.

\begin{figure}
\center
\includegraphics[width=7.5cm]{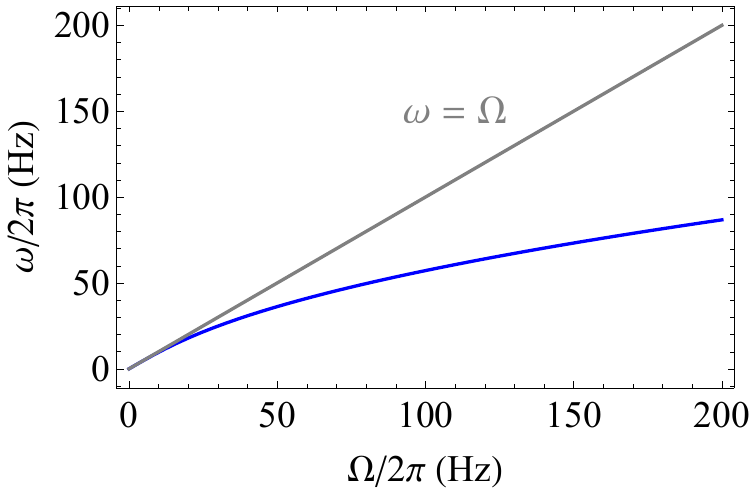}\\[0.3cm]
\includegraphics[width=7.5cm]{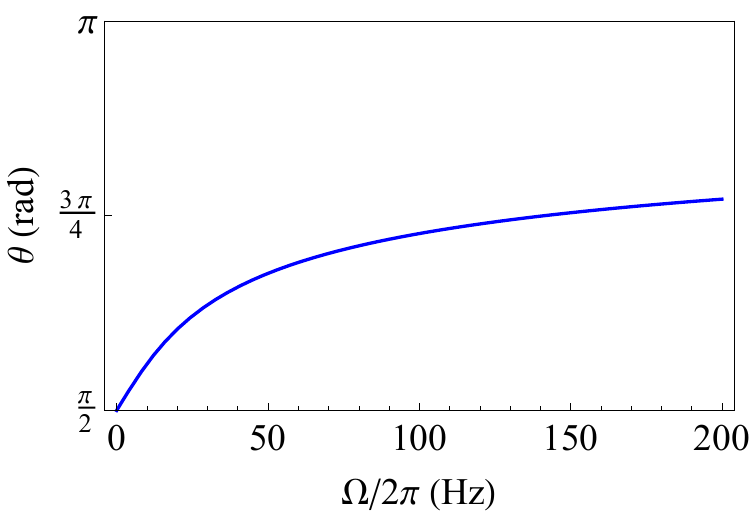}
\caption{The LFG precession frequency $\omega/(2\pi)$ (blue curve, upper plot) and the tilt angle $\theta$ (lower plot) as a function of the Larmor frequency $\Omega/(2\pi) \propto B$ for the case of $J_z=0$. The Einstein-de Haas frequency is chosen to be $\omega_I = 2\pi \times 50$\,Hz. The $\omega=\Omega$ line is shown in gray for reference in the upper plot.}
\label{fig:omega-theta-vs-Larmor-freq}
\end{figure}

Figure~\ref{fig:omega-theta-vs-Larmor-freq} shows the precession frequency $\omega$ and tilt angle $\theta$ as a function of the Larmor frequency $\Omega$ for the $J_z=0$ case.
For $B=0$, $\Omega = 0$ and $\theta = \pi/2$; this represents an LFG ``at rest'' pointing in a direction on the equator.
As $\Omega$ increases, the precession frequency $\omega$ increases as does $L_z$, and $\bs{S}$ tips below the equator ($\theta > \pi/2$) so that $S_z$ compensates the increasing $L_z$ and maintains $J_z=0$.

\begin{figure*}[t]
    \centering
    \subfloat[]{\includegraphics[width=0.32\textwidth]{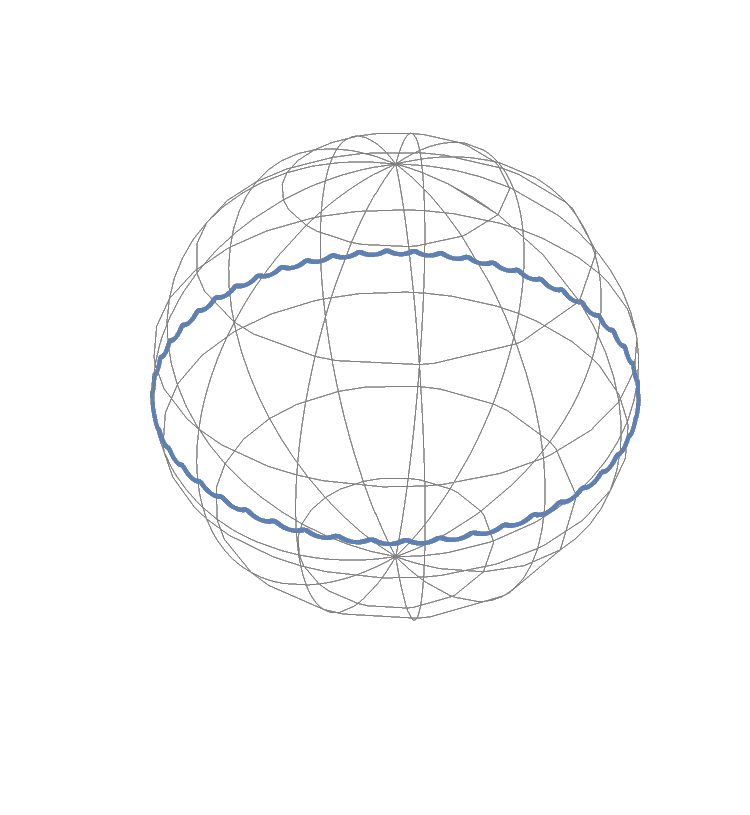}}\hfill
    \subfloat[]{\includegraphics[width=0.32\textwidth]{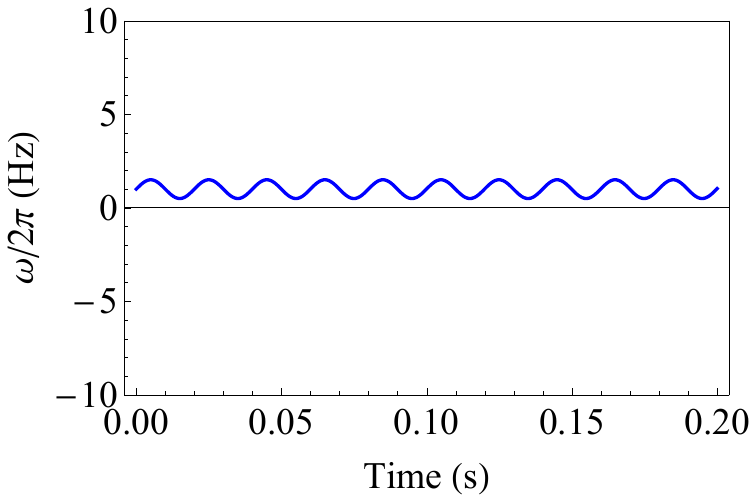}}\hfill
    \subfloat[]{\includegraphics[width=0.32\textwidth]{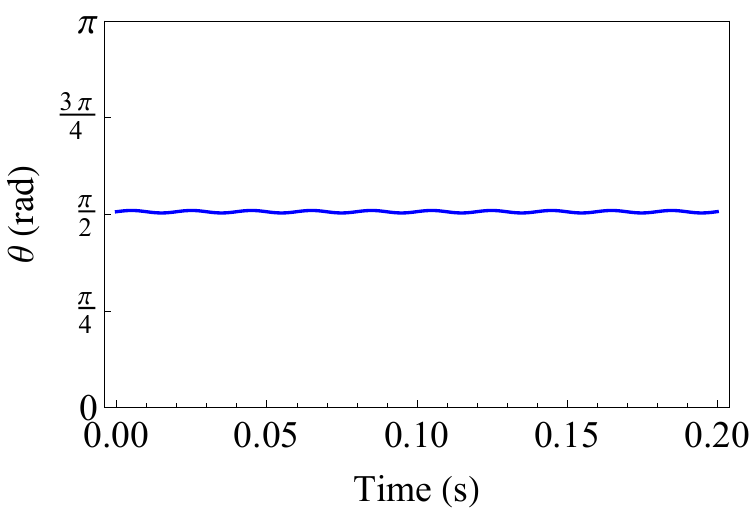}}

    \vspace{0.1cm}

    \subfloat[]{\includegraphics[width=0.32\textwidth]{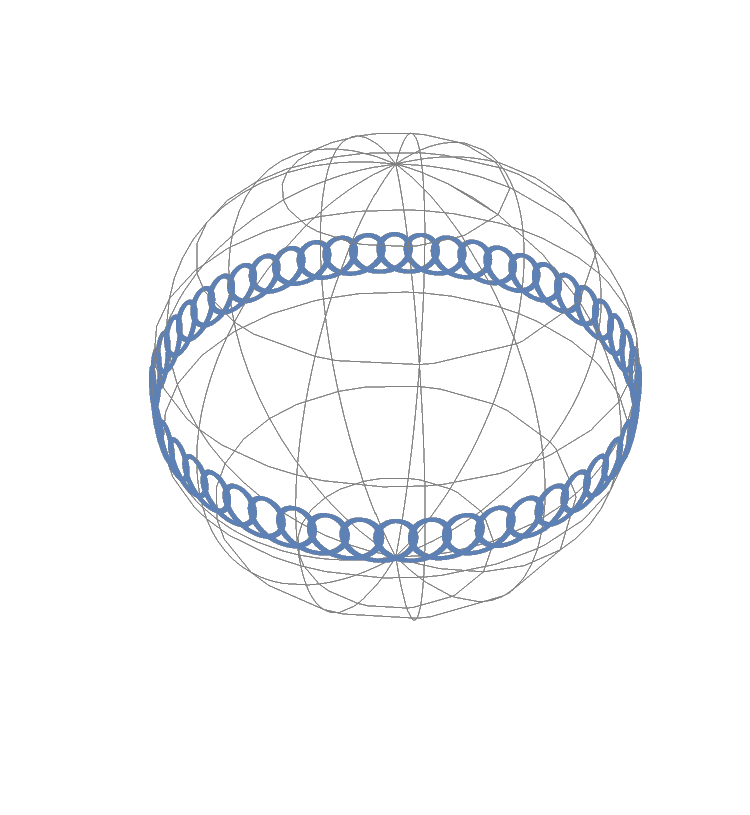}}\hfill
    \subfloat[]{\includegraphics[width=0.32\textwidth]{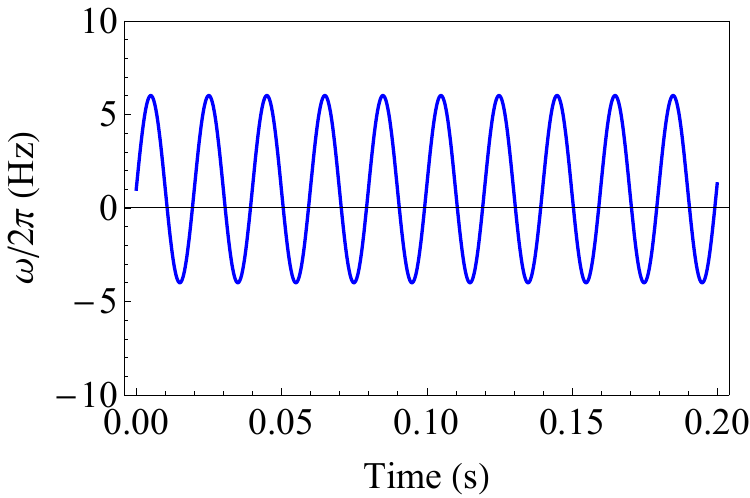}}\hfill
    \subfloat[]{\includegraphics[width=0.32\textwidth]{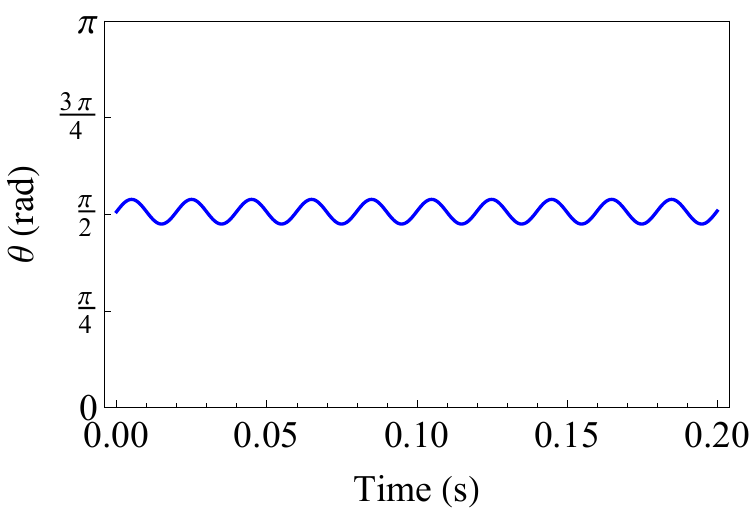}}

    \caption{Precession of an LFG with relatively small amplitude oscillations of the tilt angle $\theta$ corresponding to libration (nutation). Details of the calculations used to generate the plots are given in Appendix\,\ref{app:precession-small-amplitude-oscillations}. For all plots we choose the Einstein-de Haas frequency to be $\omega_I = 2\pi \times 50$\,Hz, the Larmor frequency to be $\Omega=2\pi \times 1$\,Hz, and the projection of the total angular momentum along $z$ (the magnetic field axis) to be zero. The upper plots [(a), (b), and (c)] correspond to a libration amplitude of $\delta\theta_0 = 0.01$\,radians and the lower plots correspond to a larger libration amplitude of $\delta\theta_0 = 0.1$\,radians. Plots (a) and (d) show the path of the unit vector $\hat{\bs{n}}$ along the direction of the spin $\bs{S}$ (see Fig.\,\ref{fig:EulerAngles}). Plots (b) and (e) show the time-dependence of the precession frequency $\omega(t)$ and plots (c) and (f) show the time-dependence of the tilt angle $\theta(t)$.}
    \label{fig:precession-with-small-oscillations}
\end{figure*}

The case of precession-dominated dynamics with nonzero but small amplitude libration is examined in Appendix\,\ref{app:precession-small-amplitude-oscillations}.
It is evident from Eq.\,\eqref{eq:Eqn-of-motion-2} and Fig.\,\ref{fig:omega-vs-theta} that libration leading to a time-varying $\theta(t)$ is necessarily coupled to a time-varying precession frequency $\omega(t)$.
We find that for libration amplitudes $\delta \theta_0 \ll 1$, the behavior of both $\theta(t)$ and $\omega(t)$ is oscillatory.
Furthermore, the oscillations of $\theta(t)$ and $\omega(t)$ are in phase with each other.
By solving Eqs.\,\eqref{eq:Eqn-of-motion-1} and \eqref{eq:Eqn-of-motion-2} in the small angle limit, we obtain expressions for the libration frequency $\omega_\ell$ and the relationship between $\delta\theta_0$ and $\delta \omega_0$, given by Eqs.\,\eqref{eq:general-form-of-libration-frequency} and \eqref{eq:relation-between-oscillation-amplitudes-of-omega-and-theta} in Appendix\,\ref{app:precession-small-amplitude-oscillations}.
The dynamical behavior of the LFG is visualized in Fig.\,\ref{fig:precession-with-small-oscillations}.
The upper plots in Fig.\,\ref{fig:precession-with-small-oscillations} show the dynamics under the condition where $\delta \omega_0 / \omega_0 \ll 1$.
The lower plots illustrate the case where $\delta \theta_0 \ll 1$ but $\delta\omega_0$ is no longer small compared to $\omega_0$.
In this case, as the LFG librates it can reverse the sense of precession, a notable phenomenon observed in classical gyroscopes (see, for example, Ref.~\cite{morin2008introduction}).

A key result of this analysis is that LFG precession-based magnetometry is not, in principle, limited only to magnetic field strengths where $\Omega \ll \omega_I$, but can, in fact, be realized for magnetic fields significantly exceeding the threshold magnetic field for precession-dominated dynamics identified in Ref.~\cite{kimball2016precessing}.

\subsection{Quantized precession near the equator ($\theta \approx \pi/2$)}
\label{subsec:equator-precession}

For an eigenstate of $J_z$ with eigenvalue $m\hbar$, the dynamics are governed by the Hamiltonian
\begin{align}
H_m\prn{ \theta,p_\theta } = \frac{p_\theta^2}{2I} + \frac{1}{2I\sin^2\theta} \prn{ m\hbar - S\cos\theta }^2 + S\Omega \cos\theta\,.
\label{eq:libration-Hamiltonian-constant-theta}
\end{align}
In this section, we consider dynamics around stable points near the equator (close to the $xy$-plane) $\abrk{\theta} = \theta_m \approx \pi/2$.
Expanding Eq.~\eqref{eq:libration-Hamiltonian-constant-theta} about small angles $\vartheta \equiv \theta - \pi/2$, and assuming that we are in the regime where the total angular momentum of the LFG is dominated by the spin $\bs{S}$, $H_m$ approximately maps to the Hamiltonian for a simple harmonic oscillator with natural frequency $\omega_I$ plus a Zeeman term (see Appendix~\ref{app:libration-constant-theta-SHO}), and thus there are quantized energy levels:
\begin{align}
E_{m,n} &\approx \hbar\omega_I \prn{n+\frac{1}{2}} + m\hbar\Omega - \frac{1}{2}I\Omega^2 \,,
\label{eq:quantized-equator-energy-levels}
\end{align}
where $n=0,1,2,\ldots$.

In this case, the expectation value of the precession frequency matches the Larmor frequency $\Omega$ \cite{kimball2016precessing}, and from Eq.~\eqref{eq:precession-op} we find that (see Appendices~\ref{app:libration-constant-theta-SHO} and \ref{app:higher-order-terms-precession-frequency})
\begin{align}
\abrk{ \hat{\omega} } \approx \frac{m\hbar}{I} \approx \Omega\,,
\label{eq:quantized-precession-frequency}
\end{align}
which shows that there are quantized ``steps'' of $\abrk{ \hat{\omega} }$ spaced by $\Delta \omega = \hbar/I$, which we define as a quantized precession frequency
\begin{align}
\Omega_Q \equiv \frac{\hbar}{I}\,.
\label{eq:quantized-precession-frequency-steps}
\end{align}
There are corresponding quantized steps in the magnetic field,
\begin{align}
\Delta B_Q = \frac{\hbar^2}{g \mu_B I} = \frac{\hbar}{g \mu_B} \Omega_Q\,.
\label{eq:quantized-magnetic-steps}
\end{align}

Note that Eq.\,\eqref{eq:quantized-precession-frequency} implies that equatorial precession requires that $m\Omega_Q \approx \Omega$, a consequence of the fact that for $\theta = \pi/2$, $S_z=0$ and therefore $J_z = m\hbar = L_z = I\Omega$.
This implies that we also have the approximate expressions for the LFG energy
\begin{align}
E &\approx \hbar\omega_I \prn{n+\frac{1}{2}} + \frac{1}{2}I\Omega^2\,, \\
&\approx \hbar\omega_I \prn{n+\frac{1}{2}} + \frac{m^2\hbar^2}{2I}\,,
\label{eq:equator-energy-levels-approximate}
\end{align}
In the case where the LFG angular momentum greatly exceeds $S$ ($J_z \gg S$) we find that the dynamics are governed by a Hamiltonian similar to that for the $J_z \ll S$ case analyzed above, but where the simple harmonic oscillator frequency is given by $m\Omega_Q \approx \Omega$ rather than $\omega_I$ (see Appendix~\ref{app:libration-constant-theta-SHO}).

Next let us consider a measurement of equatorial LFG precession.
A classical precession signal can be read out, for example, by measuring $\abrk{S_x(t)}$ using a pick-up coil positioned to measure the magnetic flux from the LFG along the $x$-axis as envisioned in Ref.~\cite{kimball2016precessing}.
However, when considering measurement of LFG precession, we must account for the fact that $\hat{J}_z$ and $\hat{\phi}$ are conjugate operators [Eq.~\eqref{eq:phi-p-phi-commutator}] corresponding to complementary observables.
An eigenstate of $\hat{J}_z$ corresponds to an LFG state with delocalized angular position $\phi$, in which case $\abrk{S_x(t)} = 0$.
Using a measurement of $\abrk{S_x(t)}$ to infer the precession frequency requires a localized wavepacket that is a superposition of many $\hat{J}_z$ eigenstates.

The semiclassical limit of the LFG dynamics is obtained when there is a narrow wave packet describing the direction of $\hat{\bs{n}}$ using the operators $\hat{\theta}$ and $\hat{\phi}$.
Let us assume that the LFG is in a coherent minimum-uncertainty wavepacket in $\prn{\phi,J_z}$ with $J \gg \hbar$ centered around the expectation value
\begin{align}
\abrk{J_z} = m_0\hbar = I\Omega = J\cos\alpha\,,
\end{align}
where $\alpha$ represents the average angle between $\bs{J}$ and $\bs{z}$.
In this coherent state, $J_z$ is a superposition over $\mc{O}\prn{\sqrt{j}}$ adjacent eigenvalues, where $j=J/\hbar$ is the total angular momentum in units of $\hbar$.
For a given expectation value $m_0$ corresponding to angle $\alpha$, the probability distribution over eigenstates of $\hat{J}_z$, with corresponding eigenvalues $J_z = m\hbar$, is given by \cite{arecchi1972atomic}
\begin{align}
P\prn{m|\alpha} = \prn{\begin{matrix}2j \\ j+m \end{matrix}} \sbrk{\cos^2\prn{\alpha/2}}^{j+m} \sbrk{\sin^2\prn{\alpha/2}}^{j-m}\,,
\label{eq:coherent-state-prob-dist}
\end{align}
where
\begin{align}
\prn{\begin{matrix} n \\ k \end{matrix}} = \frac{n!}{k!(n-k)!}
\label{eq:binomial-coefficient}
\end{align}
is the binomial coefficient.
The variance is described by
\begin{align}
\prn{\Delta m}^2 = \frac{j}{2}\sin^2\alpha\,.
\label{eq:Jz-variance}
\end{align}

Suppose that at time $t=0$ the LFG is prepared so that $\bs{S}$ is oriented along $\hat{\bs{x}}$.
In the $\Delta J_z \gg \hbar$ limit considered here, according to the central limit theorem, we can describe the initial state of the LFG [a coherent state specified by the probability distribution \eqref{eq:coherent-state-prob-dist}] as
\begin{align}
\ket{\psi(0)} = \sum_m c_m(0) \ket{j,m}\,,
\label{eq:psi0}
\end{align}
where
\begin{align}
c_m(0) \propto e^{-(m-m_0)^2/(2\Delta m)^2}
\end{align}
are the probability amplitudes for a minimum uncertainty state with $\phi(0) = 0$.
From the Hamiltonian \eqref{eq:Hamiltonian-op} in the limit $\theta \rightarrow \pi/2$, so that $\cos\theta \rightarrow 0$ and $\sin\theta \rightarrow 1$, we have
\begin{align}
\hat{H} \rightarrow \frac{\hat{p}_\theta^2}{2I} + \frac{\hat{J}_z^2}{2 I}\,,
\end{align}
therefore the energy difference between adjacent $m$-sublevels is (also see discussion in Appendix\,\ref{app:libration-constant-theta-SHO})
\begin{align}
E_m - E_{m-1} \approx \left.\pdbyd{E}{m}\right|_{m_0} \approx \frac{m_0\hbar^2}{I} = m_0\hbar\Omega_Q \approx \hbar\Omega \,.
\label{eq:Delta-Em}
\end{align}
Consequently, we have
\begin{align}
\abrk{S_x(t)}\approx \abrk{S_x(0)}\cos\prn{\Omega t}e^{-(\kappa t)^2(\Delta m)^2/2}\,
\label{eq:Sx-vs-time-equator}
\end{align}
and
\begin{align}
\kappa = \frac{1}{\hbar}\left.\pdbyd{^2E}{m^2}\right|_{m_0} = \frac{\hbar}{I} = \Omega_Q
\end{align}
governs the dephasing due to wavepacket dispersion.
In the above description we see that the precession signal comes from the phase evolution between the different $J_z$ eigenstates.
The coherence time $\tau_\phi$ of the precession signal is given by
\begin{align}
\tau_\phi \approx \frac{1}{\kappa\Delta m} \approx \frac{I}{\Delta J_z}\,.
\label{eq:coherence-time-1}
\end{align}
For a minimum uncertainty wavepacket with $\Delta\phi \ll 2\pi$,
\begin{align}
\Delta\phi \Delta J_z \approx \frac{\hbar}{2}\,,
\label{eq:DeltaPhiDeltaJz}
\end{align}
and
\begin{align}
\tau_\phi \approx \frac{2I\Delta\phi}{\hbar} = \frac{2\Delta\phi}{\Omega_Q}\,.
\label{eq:coherence-time-2}
\end{align}
This dephasing time causes an uncertainty in the measurement of precession frequency,
\begin{align}
\Delta \Omega \approx \frac{1}{\tau_\phi} \approx \frac{\Delta J_z}{I} \approx \frac{\Omega_Q}{2\Delta\phi}\,.
\label{eq:precession-freq-uncertainty}
\end{align}
The sharpness of the associated resonance can be described by the quality ($Q$) factor,
\begin{align}
Q \equiv \frac{\Omega}{\Delta\Omega} \approx 2m \Delta\phi\,
\label{eq:Q-factor-precession-near-equator}
\end{align}

For precession frequencies $\Omega \lesssim \Omega^\star = S/I = \omega_I$, the threshold for dominance of the precessional dynamics identified in Ref.\,\cite{kimball2016precessing}, $m \lesssim N/2$, and so $Q \lesssim N \Delta\phi$.
However, as noted at the beginning of this section, precessional dynamics can be observed for $\Omega > \omega_I$, and thus this is not, in fact, a limit on the attainable $Q$ for an LFG.

Equation~\eqref{eq:Q-factor-precession-near-equator} also indicates that $Q$ scales with $\Delta\phi$, suggesting that squeezing could offer advantages in quantum sensing for measurements with LFGs.
It turns out that Eq.\,\eqref{eq:DeltaPhiDeltaJz} is only valid in the limit where $\Delta\phi \ll 2\pi$ so that the small angle approximation can be used and mathematical issues related to periodicity can safely be ignored \cite{carruthers1968phase,judge1963uncertainty}.
In order to investigate the regime of large $\Delta\phi$, it is useful to employ the Robertson uncertainty relation \cite{robertson1929uncertainty},
\begin{align}
\Delta J_z \Delta\prn{ e^{i\phi} } \geq \frac{1}{2} \left| \left\langle \left[ \hat{J}_z, e^{i\hat{\phi}} \right] \right\rangle \right| = \frac{\hbar}{2} \left| \abrk{ e^{i\phi} } \right|\,.
\label{eq:Robertson-uncertainty}
\end{align}
As $\Delta \phi$ approaches the regime where the angle is completely indeterminate so that $\phi$ has a uniform distribution over all angles, $\abrk{ e^{i\phi} } \rightarrow 0$, and $\Delta J_z \rightarrow 0$.
This is the case for an eigenstate of $\hat{J}_z$.

This suggests that number-squeezed states in $J_z$ may enhance the $Q$-factor, although this comes at the cost of reduced localization in $\phi$ and hence reduced contrast in the transverse magnetization signal.
The optimal metrological tradeoff between linewidth and readout contrast, as well as alternate measurement strategies, will be explored in future work.


The above discussion clarifies that there are two complementary regimes of quantized LFG dynamics.
The semiclassical precession regime corresponds to the case where $\phi$ is localized, in which case the LFG is in a superposition of different $J_z$ states.
Alternatively, if the LFG is in an eigenstate of $\hat{J}_z$, $\phi$ is completely delocalized.
Continuous measurement of $\phi(t)$ to determine the precession frequency will tend to narrow $\Delta\phi$, consequently broadening $\Delta J_z$, shortening $\tau_\phi$ and increasing $\Delta\Omega$.

\subsection{Precession near the pole ($\theta \approx \pi$)}
\label{subsec:pole-precession}

Another interesting limiting case of the LFG dynamics is found when $\bs{S}$ is aligned along $\bs{B}$.
In this section we consider the $\theta \approx \pi$ case, which from Eq.~\eqref{eq:potential-energy} is seen to be the stable equilibrium point around the potential minimum.
Similar to our analysis of the equatorial case in Sec.\,\ref{subsec:equator-precession}, we introduce a small angle $\vartheta \equiv \theta - \pi$.
For the LFG dynamics near the pole, it is useful to distinguish between two qualitatively different cases.
The first case is when the LFG has negligible rotational angular momentum along $z$ ($L_z \approx 0$) and $\bs{S}$ passes through the pole.
For angular momentum conservation, it must be the case that $\abrk{J_z} \approx -S$ (since in our model we have assumed that the moment of inertia along the LFG axis $I_n=0$).
The second case is the more general situation where
\begin{align}
F \equiv J_z + S \neq 0\,,
\label{eq:F-definition}
\end{align}
in which case conservation of angular momentum creates an effective centrifugal barrier and the LFG instead executes small oscillations in $\vartheta$ about an equilibrium cone angle $\vartheta_0$, together with azimuthal precession.

\subsubsection{Spin passes through pole}
\label{subsec:passing-through-pole}

We begin with the special case $J_z \approx -S$, which is physically relevant, for example, if the LFG is initially prepared at rest with $\bs{S}$ pointing along $-\hat{\bs{z}}$ and then subjected to a magnetic field $\bs{B} = B\hat{\bs{z}}$.
From Eq.~\eqref{eq:precession-op}, we have
\begin{align}
\abrk{\omega} = \frac{\abrk{J_z}-S\abrk{\cos\theta}}{I\abrk{\sin^2\theta}} \approx -\omega_I\frac{ 1 + \abrk{\cos\theta} }{\abrk{\sin^2\theta}}\,.
\label{eq:precession-freq-near-pole}
\end{align}
We can rewrite the above expression in terms of the small angle $\vartheta \equiv \theta - \pi$,
\begin{align}
\abrk{\omega} \approx -\omega_I\frac{ 1 - \abrk{\cos\vartheta} }{\abrk{\sin^2\vartheta}}\,.
\label{eq:precession-freq-near-pole-varphi}
\end{align}
Expanding in $\vartheta$,
\begin{align}
\abrk{\omega} \approx -\frac{\omega_I}{2} \prn{ 1 + \frac{\abrk{\vartheta^2}}{4} + \cdots }\,.
\label{eq:precession-freq-near-pole-varphi-approximate}
\end{align}

Revisiting the Hamiltonian \eqref{eq:Hamiltonian-op} in this configuration, noting that for our new angular coordinate $\hat{p}_\vartheta = \hat{p}_\theta$ and that $J_z = p_\phi \approx -S$, we have
\begin{align}
\hat{H} & \approx \frac{1}{2I}\hat{p}_\vartheta^2 + \frac{S^2 (1 - \cos\hat{\vartheta})^2}{ 2I\sin^2\hat{ \vartheta } } - S\Omega\cos\hat{\vartheta}\,, \\
& \approx \frac{1}{2I}\hat{p}_\vartheta^2 + \frac{1}{2} S \prn{ \Omega + \frac{\omega_I}{4} } \hat{\vartheta}^2 - S\Omega\,,
\label{eq:Hamiltonian-near-pole-varphi}
\end{align}
which corresponds to a simple harmonic oscillator Hamiltonian for the librational motion with frequency
\begin{align}
\omega_\ell = \sqrt{\omega_I\Omega + \frac{\omega_I^2}{4}}\,,
\label{eq:SHO-libration-near-pole-frequency}
\end{align}
and energy eigenvalues
\begin{align}
E_n = \hbar\omega_\ell \prn{ n + \frac{1}{2} } \,,
\label{eq:SHO-libration-near-pole-energies}
\end{align}
where $n=0,1,2,\ldots$.

Note that based on Eq.\,\eqref{eq:precession-freq-near-pole-varphi-approximate}, the precession frequency depends on the expectation value of $\hat{\vartheta}^2$,
\begin{align}
\abrk{ \vartheta^2 } = \frac{\Omega_Q}{2\omega_\ell} \prn{ 2n + 1 }\,.
\label{eq:vartheta-squared-pole}
\end{align}
Of particular interest is that in the ground librational state, there are zero-point fluctuations of $\vartheta$,
\begin{align}
\left.\abrk{ \vartheta^2 }\right|_{n=0} = \frac{\Omega_Q}{2\omega_\ell} = \frac{\hbar}{2I\omega_\ell}\,,
\label{eq:ground-state-angular-spread-at-pole}
\end{align}
which, through Eq.~\eqref{eq:precession-freq-near-pole-varphi-approximate}, shift the precession frequency:
\begin{align}
\left.\abrk{\omega}\right|_{n=0} & \approx -\frac{\omega_I}{2} \prn{ 1 + \frac{\Omega_Q}{8\omega_\ell} }\,,\\
& \approx -\frac{\omega_I}{2} - \frac{\Omega_Q}{8} \frac{1}{\sqrt{ 1+ \frac{4\Omega}{\omega_I} }}\,.
\label{eq:precession-freq-near-pole-quantum-shift}
\end{align}
Near the exact south pole, the azimuthal motion of the LFG is a comparatively slow drift with frequency centered near $-\omega_I/2$.
The zero-point shift of the precession frequency is a notable purely quantum effect appearing in the LFG dynamics.

Note that Eq.~\eqref{eq:Hamiltonian-near-pole-varphi} and the subsequent analysis does not rely on the condition $\Omega \ll \omega_I$.
Rather, it follows from the small-angle expansion near the pole together with the condition $J_z \approx -S$.
Thus the harmonic libration described by Eqs.~\eqref{eq:SHO-libration-near-pole-frequency} and \eqref{eq:SHO-libration-near-pole-energies} occurs even in the high-field regime $\Omega \gg \omega_I$, provided the rigid-macrospin description remains valid.

\subsubsection{General LFG polar dynamics}
\label{subsec:general-polar-dynamics}

In the more general near-pole case where $F = J_z + S \neq 0$, the pole itself is no longer the minimum of the effective potential.
Instead, angular-momentum conservation produces a centrifugal barrier that tilts the LFG away from $\theta = \pi$.
Expanding in the small angle $\vartheta \equiv \theta - \pi$, we find
\begin{align}
J_z - S\cos\theta \approx \prn{J_z + S} - \frac{S}{2}\vartheta^2 = F - \frac{S}{2}\vartheta^2\,,
\end{align}
which can be substituted into Eq.\,\eqref{eq:Hamiltonian-op} yielding
\begin{align}
\hat{H} \approx \frac{\hat{p}_\vartheta^2}{2I} + \frac{F^2}{2I \hat{\vartheta}^2} + \frac{1}{2}S\prn{ \Omega + \frac{\omega_I}{4} }\hat{\vartheta}^2 - \frac{\omega_I}{2}F-S\Omega\,.
\label{eq:approximate-Hamiltonian-general-pole}
\end{align}
The corresponding effective potential is
\begin{align}
U\ts{eff}(\vartheta) = \frac{F^2}{2I\vartheta^2} + \frac{1}{2}S \prn{\Omega + \frac{\omega_I}{4}} \vartheta^2 + \text{const.}
\label{eq:polar_cone_potential}
\end{align}
The first term in Eq.~\eqref{eq:polar_cone_potential} is a centrifugal barrier.
Minimizing the effective potential for $\vartheta$,
\begin{align}
\left.\frac{dU_{\rm eff}}{d\vartheta}\right|_{\vartheta = \vartheta_0} = -\frac{F^2}{I\vartheta_0^3} + S\prn{\Omega + \frac{\omega_I}{4}}\vartheta_0 = 0\,,
\end{align}
we can solve for $\vartheta_0$,
\begin{align}
\vartheta_0^2 = \frac{\left| F \right|}{I\omega_\ell}\,,
\label{eq:vartheta0-general-pole}
\end{align}
where $\omega_\ell$ is given by Eq.\,\eqref{eq:SHO-libration-near-pole-frequency}.
The frequency of small oscillations about $\vartheta_0$ is given by
\begin{align}
\omega_\vartheta^2 = \frac{1}{I} \left.\frac{d^2U_{\rm eff}}{d\vartheta^2}\right|_{\vartheta = \vartheta_0} = 4 \omega_\ell^2\,,
\label{eq:librational-oscillation-freq-near-pole}
\end{align}
and the precession frequency based on Eq.~\eqref{eq:precession-op} is
\begin{align}
\omega \approx \frac{F}{I\vartheta_0^2}-\frac{\omega_I}{2} \approx \pm \omega_\ell - \frac{\omega_I}{2} \,.
\label{eq:precession-freq-near-south-pole-general}
\end{align}

In the high-field regime $\Omega \gg \omega_I$, the librational oscillation frequency about $\vartheta_0$ is
\begin{align}
\omega_\vartheta \approx 2 \sqrt{\Omega \omega_I}\,,
\end{align}
and the precession frequency is
\begin{align}
\omega \approx \pm \sqrt{\Omega \omega_I}\,.
\end{align}
Therefore, in the general case, we find that both the libration and precession frequencies are $B$-dependent for $\Omega \gg \omega_I$, in contrast to the ``exact-pole'' case where the precession is approximately $B$-independent for $\Omega \gg \omega_I$ [Eq.\,\eqref{eq:precession-freq-near-pole-quantum-shift}].

For the purposes of cooling and state preparation discussed in Sec.\,\ref{sec:cooling}, the exact-pole branch is a natural starting point since it corresponds to an LFG prepared at rest with $\bs{\mu}$ along $\bs{B}$.
However, small-angle polar dynamics are far more general in practice than the special condition $J_z=-S$, and thus precession is possible in the polar geometry for $\Omega \gg \omega_I$, as was found for the equatorial case in Sec.\,\ref{subsec:equator-precession}.

\subsection{Ladder spectroscopy of quantized precession and libration levels}
\label{subsec:ladder-spectroscopy}

We have seen that the quantum dynamics of the LFG are related to a ladder of different eigenstates of $\hat{J}_z$ and a ladder of different vibrational levels related to librational motion.
This naturally leads to the question of how such quantum states might be observed and how they might be useful for metrology.
Note that Refs.\,\cite{kimball2016precessing,fadeev2021ferromagnetic} implicitly consider \emph{coherent} states of the LFG in minimum-uncertainty states of the angular position observables $\hat{\theta}$ and $\hat{\phi}$.
As discussed in Sec.\,\ref{subsec:equator-precession}, such LFG states are superpositions of a relatively large number of $J_z$ eigenstates.
In the present work we have identified a qualitatively different regime where the LFG is in an eigenstate of $\hat{J}_z$ and a delocalized state of the angular observable $\hat{\phi}$.
In this section we describe how a radio-frequency (rf) field can be used, in principle, to drive transitions between these quantized levels and how such ladder spectroscopy could be implemented experimentally.


\subsubsection{Quantized energy levels near the equator ($\theta \approx \pi/2$)}
\label{subsubsec:quantized-energy-levels-equator}

\begin{figure}
\center
\includegraphics[width=8.6cm]{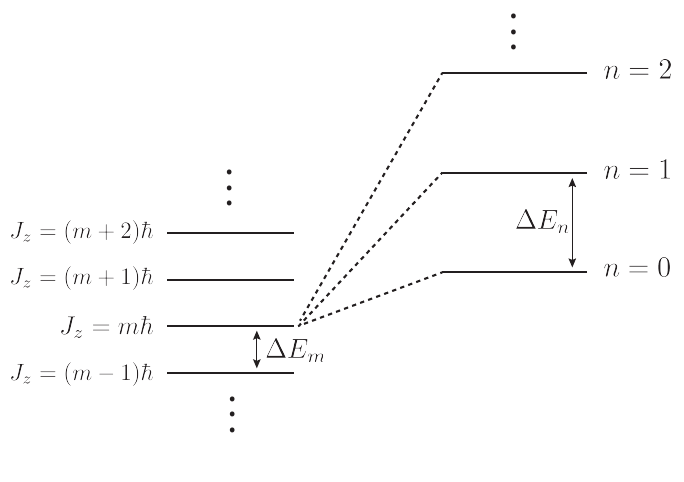}
\caption{Energy level diagram showing the quantized states of $J_z$ and simple harmonic motion associated with libration near the equator ($\theta \approx \pi/2$). The different $J_z=m\hbar$ levels are separated in energy by $\Delta E_m \approx \hbar\Omega$ and the librational energy levels are separated in energy by $\Delta E_n$ as described by Eqs.\,\eqref{eq:delta-En-equator} and \eqref{eq:delta-En-equator-large-Jz}.}
\label{fig:equator-energy-levels}
\end{figure}

For the equatorial case of $\abrk{\theta} = \theta_m \approx \pi/2$, the energies of the quantum states of the LFG are given by Eq.\,\eqref{eq:quantized-equator-energy-levels}, shown schematically in Fig.\,\ref{fig:equator-energy-levels}.
The energy difference between adjacent $m$-levels holding $n$ constant is
\begin{align}
\Delta E_m \equiv E_{m+1,n} - E_{m,n} \approx \hbar\Omega\,.
\label{eq:delta-Em-equator}
\end{align}
For the case where $J_z \ll S$ ($m \ll N/2$), the energy difference between adjacent $n$-levels holding $m$ constant is
\begin{align}
\Delta E_n \equiv E_{m,n+1} - E_{m,n} \approx \hbar\omega_I \approx \frac{N}{2} \frac{\hbar^2}{I}\,,
\label{eq:delta-En-equator}
\end{align}
while for the case where $J_z \gg S$ ($m \gg N/2$),
\begin{align}
\Delta E_n \approx \hbar m \Omega_Q \approx m \frac{\hbar^2}{I}\,
\label{eq:delta-En-equator-large-Jz}
\end{align}
and, furthermore, $\Delta E_n \approx \hbar\Omega$ due to the equatorial precession frequency condition \eqref{eq:quantized-precession-frequency}.

\subsubsection{Quantized energy levels near the south pole ($\theta \approx \pi$)}
\label{subsubsec:quantized-energy-levels-pole}

Near the south pole, the value of the potential at the minimum $\vartheta = \vartheta_0$ based on Eqs.\,\eqref{eq:polar_cone_potential} and \eqref{eq:vartheta0-general-pole} is
\begin{align}
\left.U_{\rm eff}\right|_{\vartheta = \vartheta_0} = \left| F \right| \sqrt{ \Omega\omega_I + \frac{\omega_I^2}{4} } - \frac{\omega_I}{2}F - S\Omega\,,
\end{align}
and so the LFG energy, including the librational oscillation, is
\begin{align}
E = \left| F \right| \sqrt{ \Omega\omega_I + \frac{\omega_I^2}{4} } - \frac{\omega_I}{2}F - S\Omega + \hbar\omega_\vartheta \prn{ n + \frac{1}{2} }\,.
\end{align}
The energy difference between adjacent $m$-levels holding $n$ constant is found by noting that
\begin{align}
\pdbyd{E}{m} = \pdbyd{E}{F}\pdbyd{F}{m} = \hbar\pdbyd{E}{F}\,,
\end{align}
and we find that\footnote{Equation~\eqref{eq:delta-Em-near-pole} is derived from $$\Delta E_m \approx \prn{\left| F + \hbar \right| - \left| F \right|} \sqrt{ \Omega\omega_I + \frac{\omega_I^2}{4} } - \frac{\hbar\omega_I}{2}\,.$$}
\begin{align}
\Delta E_m \approx \pm\hbar\sqrt{ \Omega\omega_I + \frac{\omega_I^2}{4} } - \frac{\hbar\omega_I}{2}\,.
\label{eq:delta-Em-near-pole}
\end{align}
For the exact pole case, there is no $m$-dependence of the energy levels since $J_z = -S$ is fixed.

The vibrational ladder can arise either from libration about the exact pole for $J_z = -S$ or from small oscillations about a finite cone angle for $J_z + S \neq 0$.
In the exact pole case,
\begin{align}
\Delta E_n \approx \hbar\omega_\ell = \hbar \sqrt{ \Omega \omega_I + \frac{\omega_I^2}{4} }\,,
\label{eq:delta-En-pole-exact}
\end{align}
and in the near-pole case,
\begin{align}
\Delta E_n \approx 2 \hbar\omega_\ell  \,.
\label{eq:delta-En-pole-near}
\end{align}

\subsubsection{Perturbing Hamiltonian from an rf-drive}
\label{subsubsec:rf-drive-hamiltonian}

If we add a relatively weak time-dependent rf field $\bs{B}\ts{rf}(t)$ to the existing static field $\bs{B}$, we introduce an additional term into the Hamiltonian \eqref{eq:Hamiltonian}:
\begin{align}
\mc{V}\ts{rf}(t) = \frac{g\mu_B}{\hbar} \bs{S}\cdot\bs{B}\ts{rf}(t) = \frac{g\mu_B}{\hbar} S \hat{\bs{n}}\cdot\bs{B}\ts{rf}(t)\,.
\label{eq:rf-potential-general}
\end{align}

Near the equator ($\theta \approx \pi/2$), the LFG orientation can be characterized in the lab frame $(x,y,z)$ as pointing along $\hat{\bs{n}} \approx (\cos\phi,\sin\phi,0)$.
We choose a circularly polarized rf field propagating along $z$,
\begin{align}
\bs{B}\ts{rf}(t) = B_1\sbrk{ \hat{\bs{x}} \cos\prn{\omega\ts{rf}t} + \hat{\bs{y}} \sin\prn{\omega\ts{rf}t} }\,,
\end{align}
so that
\begin{align}
\hat{\bs{n}}\cdot\bs{B}\ts{rf}(t) & = B_1 \sbrk{ \cos\phi\cos\prn{\omega\ts{rf}t} + \sin\phi\sin\prn{\omega\ts{rf}t} }\,, \\
& = B_1 \cos\prn{ \phi - \omega\ts{rf}t }\,,
\end{align}
and consequently
\begin{align}
\hat{V}\ts{rf}(t) &= \frac{g\mu_B}{\hbar} S B_1 \cos\prn{ \hat{\phi} - \omega\ts{rf}t }\,, \\
& = \frac{g\mu_B S B_1}{2\hbar}\prn{ e^{i(\hat{\phi} - \omega\ts{rf}t)} + e^{-i(\hat{\phi} - \omega\ts{rf}t)}}\,.
\label{eq:rf-potential-circular}
\end{align}
Since the eigenstates of $\hat{J}_z$, $\ket{m}$, satisfy $\langle \phi \ket{m} \propto e^{im\phi}$, we obtain the standard selection rule
\begin{align}
\bra{m'} e^{\pm i\hat{\phi}} \ket{m} \propto \delta_{m',m \pm 1}\,.
\end{align}
Thus $\hat{V}\ts{rf}(t)$ couples $\ket{m}$ to $\ket{m \pm 1}$ states, and drives $\Delta m = \pm 1$ transitions.\footnote{It turns out that this approach works for any angle $\theta$, but the math is somewhat simplified near the equator.}

At the south pole ($\theta = \pi$), we have the Hamiltonian \eqref{eq:Hamiltonian-near-pole-varphi} written in terms of the small angle $\vartheta \equiv \theta - \pi$, and librational eigenstates $\ket{n}$ with energies given by Eq.\,\eqref{eq:SHO-libration-near-pole-energies}; near but not exactly at the south pole, we have a similar Hamiltonian but with the oscillator frequency $\omega_\vartheta$ given by Eq.\,\eqref{eq:librational-oscillation-freq-near-pole} and the oscillations centered about $\vartheta_0$.
The LFG orientation is given by $\hat{\bs{n}} \approx (-\vartheta\cos\phi,-\vartheta\sin\phi,1-\vartheta^2/2)$.
If the weak rf field is applied along $\bs{x}$,
\begin{align}
\bs{B}\ts{rf}(t) = B_1 \hat{\bs{x}} \cos\prn{\omega\ts{rf}t}\,,
\end{align}
and
\begin{align}
\hat{\bs{n}}\cdot\bs{B}\ts{rf}(t) = -B_1 \vartheta \cos\phi \cos\prn{\omega\ts{rf}t}\,.
\end{align}
For libration initially in the $xz$ plane, so that the initial angle $\phi \approx 0$ and $\cos\phi \approx 1$,
\begin{align}
\hat{V}\ts{rf}(t) \approx -\frac{g\mu_B}{\hbar} S B_1 \hat{\vartheta} \cos\prn{\omega\ts{rf}t}\,.
\end{align}
The operator $\hat{\vartheta}$ can be written in terms of the simple harmonic oscillator raising and lowering operators, $\hat{a}^\dagger$ and $\hat{a}$, respectively:
\begin{align}
\hat{\vartheta} = \sqrt{ \frac{\hbar}{2 I \omega_\ell} } \prn{ \hat{a}^\dagger + \hat{a} }\,,
\end{align}
so
\begin{align}
\hat{V}\ts{rf}(t) \approx -\frac{g\mu_B}{\hbar} S B_1 \sqrt{ \frac{\hbar}{2 I \omega_\ell} } \prn{ \hat{a}^\dagger + \hat{a} } \cos\prn{\omega\ts{rf}t}\,.
\end{align}
The matrix elements $\bra{n'} \prn{ \hat{a}^\dagger + \hat{a} } \ket{n}$ are nonzero only for $\Delta n = \pm 1$, so $\hat{V}\ts{rf}(t)$ will drive transitions between neighboring librational levels.\footnote{It turns out that this approach also works for any angle $\theta$.}

For the near-pole case, the same physics applies, but $\vartheta \rightarrow \delta\hat{\vartheta} = \vartheta - \vartheta_0$ and the oscillator frequency $\omega_\ell \rightarrow \omega_\vartheta$, see Eq.\,\eqref{eq:librational-oscillation-freq-near-pole}.

\subsubsection{Observables}
\label{subsubsec:rf-drive-observables}

The dynamics of the LFG can be observed by measuring the field from the magnetization $\bs{\mu} = -g\mu_B\bs{S}/\hbar$.
An intuitive way to observe precession in the equatorial case, as originally envisioned in Ref.\,\cite{kimball2016precessing}, is by measuring the time-dependent transverse magnetization with a pickup loop along, e.g., $x$, to determine
\begin{align}
\abrk{S_x\prn{t}} \approx S\cos\phi\prn{t}\,.
\end{align}
This is the approach considered in Sec.\,\ref{subsec:equator-precession} and the measurable time-dependent signal is given by Eq.\,\eqref{eq:Sx-vs-time-equator}.

Another possibility is to measure the $z$-magnetization using, e.g., the magnetic flux through a pickup loop along $z$ in order to read out the $z$-component of the macrospin $\bs{S}$,
\begin{align}
S_z = S \cos\theta\,.
\end{align}
For the $\omega = \Omega$ case the vertical magnetization crosses zero when $m\hbar = I\Omega$, which occurs at discrete, quantized field values [Eq.\,\eqref{eq:quantized-magnetic-steps}],
\begin{align}
B_m = \frac{m \hbar \Omega_Q}{g \mu_B} = \frac{m \hbar^2}{g \mu_B I}\,.
\label{eq:quantized-fields-equator}
\end{align}
These discrete field values are a signature of $J_z$ quantization for equatorial precession.
Note that observing $S_z$ would constitute a back-action-evading measurement of the precession, as it is insensitive to the phase $\phi$ of the LFG.

Exactly at the south pole ($\theta = \pi$), $\abrk{J_z} = -S$ and $\abrk{S_z} \approx -S + \mc{O}(\vartheta^2)$ (where $\vartheta = \theta-\pi$), so the measurement of vertical magnetization is relatively insensitive to the librational motion.
Instead, if the transverse magnetization is measured with a pickup loop along, e.g., $x$,
\begin{align}
\abrk{S_x\prn{t}} \approx -S \vartheta\prn{t} \cos\phi\prn{t} \,,
\label{eq:south-pole-spin-measurement}
\end{align}
oscillations of $\vartheta\prn{t}$ at $\omega_\ell$ can be directly observed as well as precession at $\omega \approx -\omega_I/2$ as described by Eq.\,\eqref{eq:precession-freq-near-pole-quantum-shift}.

Near but not exactly at the south pole, the nonzero $\vartheta_0$ gives a small static offset value for a transverse magnetization measurement, with precession frequency given by Eq.\,\eqref{eq:precession-freq-near-south-pole-general} and a librational oscillation frequency described by Eq.\,\eqref{eq:librational-oscillation-freq-near-pole}.

\subsubsection{Spectroscopy of the $J_z$ ladder at the equator}
\label{subsubsec:Jz-spectroscopy}

In the case of precession-dominated dynamics near the equator, driving transitions between $m$-sublevels using a circularly polarized rf field $\bs{B}\ts{rf}(t)$ as described in Sec.\,\ref{subsubsec:rf-drive-hamiltonian} will change the values of the observables described above in Sec.\,\ref{subsubsec:rf-drive-observables}.
The LFG precession frequency is calculated to higher order in $m$ in Appendix\,\ref{app:higher-order-terms-precession-frequency} [Eq.\,\eqref{eq:omega_mn}], and the shift induced when driving $m \rightarrow m \pm 1$ transitions is given by
\begin{align}
\Delta \omega_{m} &= \pm\pdbyd{\omega_{mn}}{m} \nonumber \\
&\approx \pm\frac{\Omega^2}{\omega_I^2}\Omega_Q \pm \frac{\Omega_Q^2}{2\omega_I} \prn{ 2n+1 } \pm \frac{6m\Omega_Q^2}{\omega_I} \prn{ m\Omega_Q - \Omega }\,.
\label{eq:precession-freq-shift-with-Delta-m}
\end{align}
Since $\Omega \approx m\Omega_Q$ near the equator (Appendix\,\ref{app:libration-constant-theta-SHO}), the last term can be neglected and we have
\begin{align}
\Delta \omega_{m} \approx \pm\frac{\Omega^2}{\omega_I^2}\Omega_Q \pm \frac{\Omega_Q^2}{2\omega_I} \prn{ 2n+1 }\,.
\label{eq:Delta-omega_m}
\end{align}
For $n \ll m$, the second term in Eq.\,\eqref{eq:Delta-omega_m} is negligible.

For the spin projection along $z$ [Eq.\,\eqref{eq:S_z_mn}], driving $m \rightarrow m \pm 1$ transitions changes $\abrk{S_z}$ by
\begin{align}
\Delta \abrk{S_z}_m &= \pm \pdbyd{S_z}{m} \nonumber \\
&\approx \pm S \frac{\Omega_Q}{\omega_I}\,.
\end{align}
In terms of the measurable change in magnetization along $z$,
\begin{align}
\Delta \abrk{\mu_z}_m = \mp \frac{g\mu_B}{\hbar}\Delta\abrk{S_z}_m \approx \mp g \mu_B \frac{S}{\hbar} \frac{\Omega_Q}{\omega_I} = \mp g \mu_B\,.
\end{align}

From Eqs.\,\eqref{eq:delta-Em-equator} and \eqref{eq:rf-potential-circular}, we can see that there is a resonant condition achieved when $\omega\ts{rf} = \Omega$.
For a fixed magnetic field $B$, with the LFG in an initial state $\ket{m,n}$, sweeping $\omega\ts{rf}$ would reveal a resonance in the response of $\abrk{\mu_z}$ and a shift of the precession frequency $\Delta \omega_m$.

If $\omega\ts{rf}$ is fixed, with the LFG in an initial state $\ket{m,n}$, and the magnetic field $B$ is swept, the precession quantization condition [Eqs.\,\eqref{eq:quantized-precession-frequency} and \eqref{eq:quantized-magnetic-steps}] implies that for the given $m$ there is a special field $B_m$ [Eq.\,\eqref{eq:quantized-fields-equator}] for which $\abrk{L_z} = I\Omega = m\hbar$ exactly matches $\abrk{J_z}$ such that $\abrk{S_z}_m = \abrk{\mu_z}_m = 0$.
Near $B_m$,
\begin{align}
\abrk{S_z}_m = m\hbar - I\Omega = \frac{g \mu_B I}{\hbar} \prn{ B_m - B }\,;
\end{align}
the vertical magnetization has a distinct zero-crossing at each $B_m$ value.
The quantized zero-crossings of the vertical magnetization are a signature of the quantized precession:
as $B$ is tuned, the condition $I\Omega = m\hbar$ can only be satisfied at discrete field values.
This is analogous to flux quantization in Superconducting Quantum Interference Devices (SQUIDs) \cite{clarke2006squid}.

\subsubsection{Spectroscopy of polar librational levels}
\label{subsubsec:SHO-spectroscopy}

Likewise, as discussed in Sec.\,\ref{subsubsec:rf-drive-hamiltonian}, a linearly polarized $\bs{B}\ts{rf}(t)$ can be used to drive $n \rightarrow n \pm 1$ transitions between librational levels.
Near the south pole ($\theta \approx \pi$), the resonant condition is achieved when
\begin{align}
\omega\ts{rf} = \omega_\vartheta = 2\omega_\ell = 2 \sqrt{ \Omega \omega_I + \frac{\omega_I^2}{4} }\,,
\end{align}
and an increase in the amplitude of the librational oscillations can be detected [Eq.\,\eqref{eq:south-pole-spin-measurement}].
Scanning the magnetic field with a fixed $\omega\ts{rf}$, the libration will exhibit a resonance at
\begin{align}
B = \frac{\hbar}{4g\mu_B} \prn{\frac{\omega\ts{rf}^2}{\omega_I} - \omega_I}\,.
\end{align}
(For the exact pole case, the resonant condition is $\omega\ts{rf} = \omega_\ell$.)

Generally, rf-spectroscopy of $m$- and $n$-levels can be carried out for any $\theta$.


\section{Magnetic resonance}
\label{sec:magnetic-resonance}

In Sec.\,\ref{subsubsec:rf-drive-hamiltonian} we showed that a circularly polarized rf drive field can induce transitions between $m$-levels (eigenstates of $\hat{J}_z$) and a linearly polarized rf drive field can induce transitions between $n$-levels (eigenstates of the librational harmonic motion).
Therefore rf fields can be used to manipulate the LFG dynamics as well as carry out spectroscopic measurements, opening a new tool for experiments with LFGs.
Furthermore, we have seen throughout Sec.\,\ref{sec:precession-small-libration} that the precession and libration/nutation dynamics, as well as the tilt angle $\theta$, are affected by the applied magnetic field.
In combination, these observations imply that techniques related to the magnetic resonance methods applied in experiments with electron and nuclear spins in atoms, molecules, and condensed matter systems can be used to control and measure LFG dynamics.
Note that the resonances described in this section have both a classical interpretation, as enhanced response of the coupled precession-libration dynamics, and a quantum interpretation, as rf-driven transitions between discrete $m$- and $n$-levels.

\begin{figure*}
    \centering
    \subfloat[]{\includegraphics[width=0.32\textwidth]{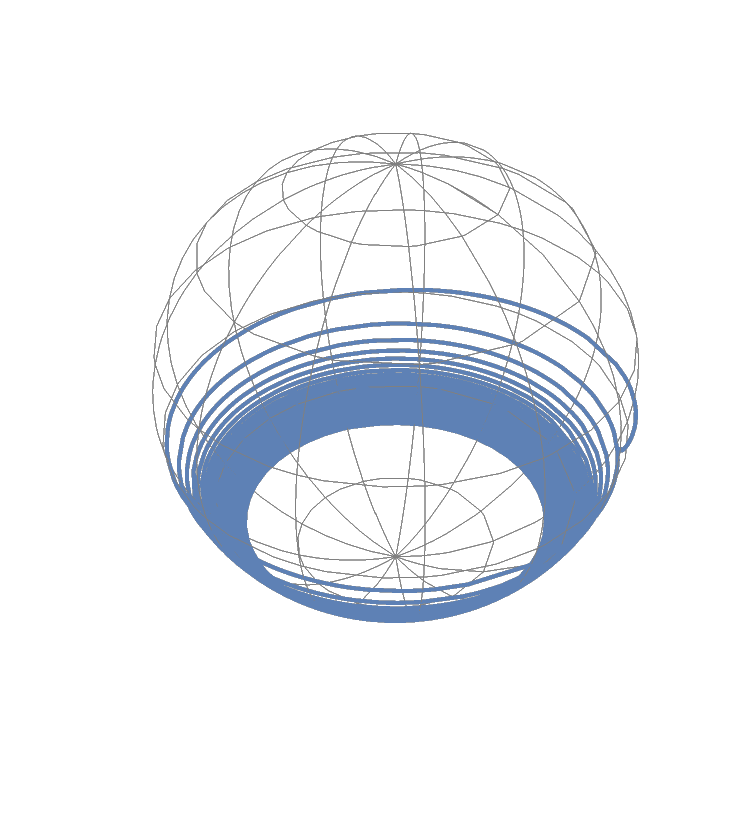}}\hfill
    \subfloat[]{\includegraphics[width=0.32\textwidth]{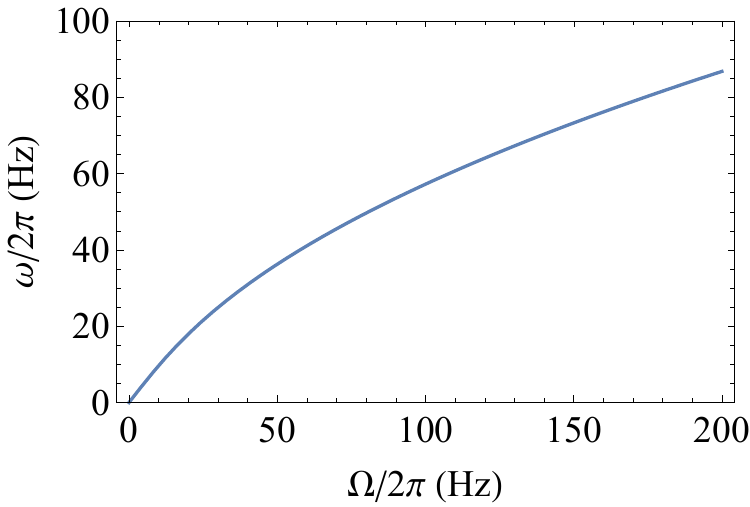}}\hfill
    \subfloat[]{\includegraphics[width=0.32\textwidth]{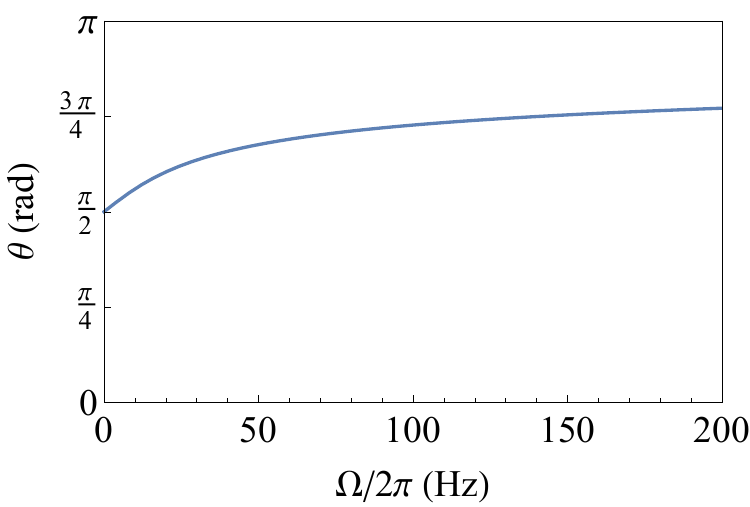}}

    \vspace{0.1cm}

    \subfloat[]{\includegraphics[width=0.32\textwidth]{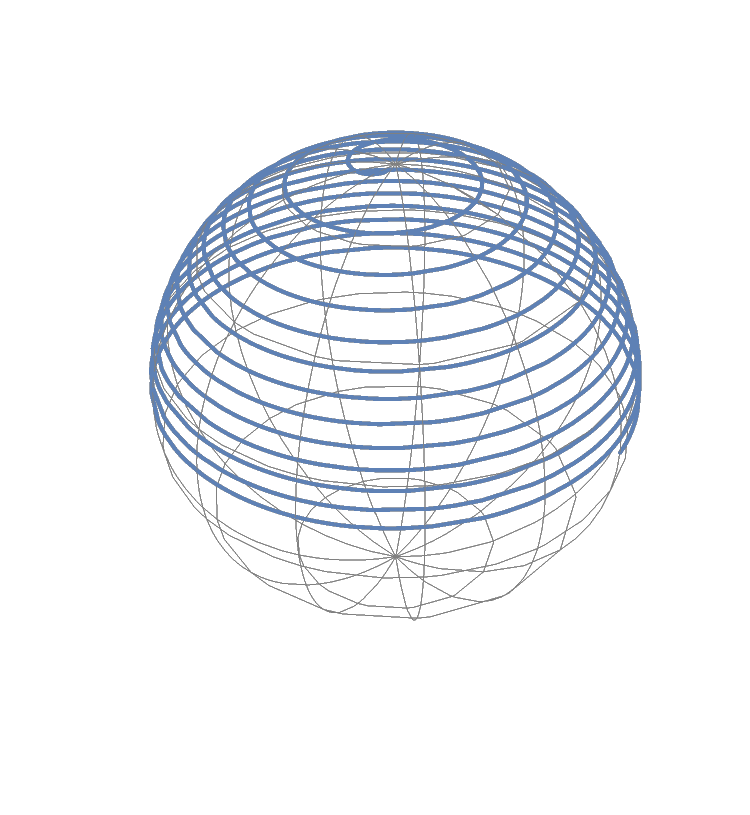}}\hfill
    \subfloat[]{\includegraphics[width=0.32\textwidth]{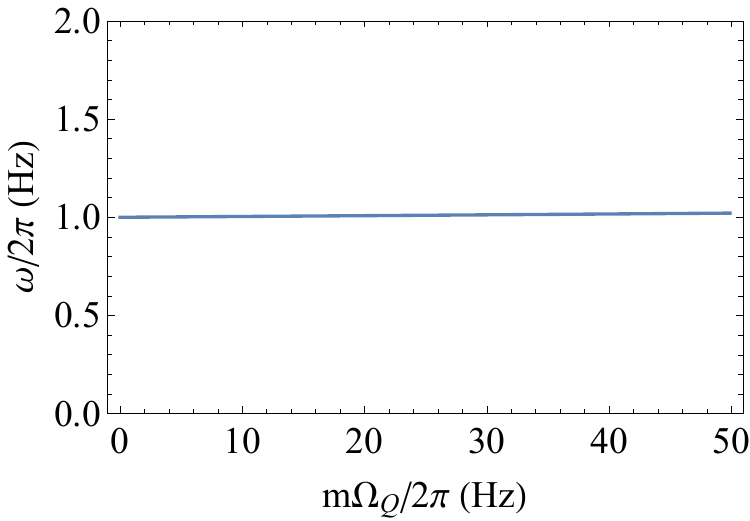}}\hfill
    \subfloat[]{\includegraphics[width=0.32\textwidth]{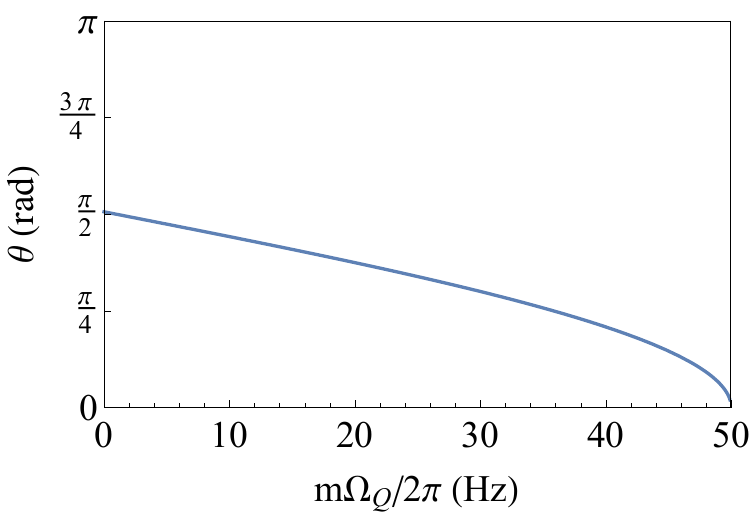}}

    \caption{Upper plots (a), (b), and (c) show the LFG dynamical behavior as a function of Larmor frequency $\Omega$, proportional to the magnetic field $B$, for $J_z=0$. Lower plots (d), (e), and (f) show the LFG dynamical behavior as a function of $m\Omega_Q$, proportional to $J_z$, for a Larmor frequency of $\Omega = 2\pi \times 1$\,Hz.  For all plots we choose the Einstein-de Haas frequency to be $\omega_I = 2\pi \times 50$\,Hz and neglect librational/nutational motion. As in Fig.\,\ref{fig:precession-with-small-oscillations}, plots (a) and (d) show the angular position on the unit sphere of the vector $\hat{\bs{n}}$ (along the direction of the spin $\bs{S}$, see Fig.\,\ref{fig:EulerAngles}); plots (b) and (e) show the precession frequency $\omega$; plots (c) and (f) show the tilt angle $\theta$. We assume a linear in time ramp of $\Omega$ for plot (a) and $m\Omega_Q$ for plot (d), respectively, at a rate of 1\,Hz/s, and the precession frequency is artificially reduced by a factor of 400 in plots (a) and (d) for illustrative purposes. Angular momentum $J_z$ can be manipulated by driving the LFG with a circularly polarized rf field (Sec.\,\ref{subsubsec:rf-drive-hamiltonian}), thereby increasing (decreasing) the quantum number $m$. Note from plot (e) that the precession frequency $\omega$ is essentially constant as $m$ changes and $\omega \approx \Omega$, indicating that under the illustrated conditions, tuning the rf drive frequency to $\Omega$ yields a strong response in $\theta$.}
    \label{fig:LFG-dynamics-rf-drive-magnetic-field-scan}
\end{figure*}

Figure\,\ref{fig:LFG-dynamics-rf-drive-magnetic-field-scan} illustrates the dynamical response of an LFG to magnetic field changes and an rf drive field.
The resulting dynamics are calculated by simultaneously solving Eqs.\,\eqref{eq:p-phi} and \eqref{eq:quadratic-eq-for-precession-freq}.
Libration (nutation) is ignored in Fig.\,\ref{fig:LFG-dynamics-rf-drive-magnetic-field-scan}.
The upper plots, (a) -- (c), illustrate how ramping the magnetic field (with the LFG starting from rest pointing along the equator, $\theta = \pi/2$) increases the precession frequency $\omega$ and tilts the LFG away from the equator.
In Fig.\,\ref{fig:LFG-dynamics-rf-drive-magnetic-field-scan}(c), as well as in Fig.\,\ref{fig:omega-theta-vs-Larmor-freq}, we observe that for $\Omega \ll \omega_I$, $\theta$ increases linearly as a function of $\Omega$ whereas for $\Omega \gg \omega_I$, $\theta$ increases much more slowly as a function of $\Omega$.
This can be understood by combining Eqs.\,\eqref{eq:quadratic-eq-for-precession-freq} and \eqref{eq:prec-freq-vs-angle-Jz-is-zero}, from which we find
\begin{align}
\frac{\Omega}{\omega_I} = - \frac{\cos\theta}{\sin^4\theta}\,.
\end{align}
Starting from $\theta = \pi/2$ and $\Omega = 0$, $\theta$ will increase linearly as $\Omega$ increases for $\left|\cos\theta\right| \ll 1$.
However, when $\Omega/\omega_I \gg 1$, and $\theta \rightarrow \pi$, the tilt angle as a function of Larmor frequency scales as $\theta \sim \pi - \prn{ \Omega/\omega_I }^{-1/4}$.
The lower plots, (d) -- (f), illustrate how driving $m \rightarrow m+1$ transitions with an rf field, thereby increasing $m\Omega_Q$ and, proportionally, $J_z$, affect the LFG.
Notably, Fig.\,\ref{fig:LFG-dynamics-rf-drive-magnetic-field-scan}(e) indicates that the precession frequency is mostly independent of $m$, $\omega \approx \Omega$, as is indicated also by Eq.\,\eqref{eq:Delta-omega_m}, which shows that for the chosen parameters $\Delta \omega_m \ll \omega$.
This shows that, under these conditions, a nearly constant frequency rf field matching the precession frequency can drive many successive $m \rightarrow m \pm 1$ transitions and be used to control the LFG tilt angle.
Furthermore, one expects a narrow magnetic resonance near $\omega\ts{rf} \approx \omega \approx \Omega$, with the resonance width determined by relaxation, dephasing, and power-broadening mechanisms.
Since the relevant dissipation rates can in principle be made much smaller than $\Omega$ \cite{kimball2016precessing,fadeev2021ferromagnetic}, this resonance can be extremely narrow, and offer a useful technique for magnetic field measurements with LFGs.

\begin{figure}
    \includegraphics[width=7.5cm]{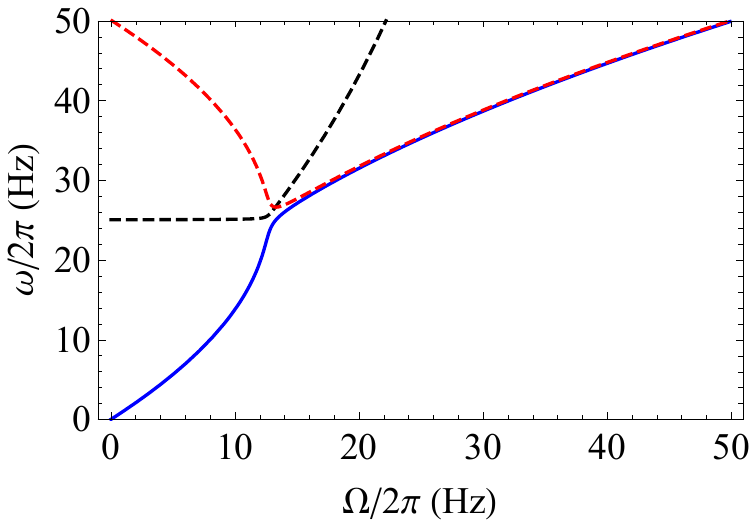}\\[0.3cm]
    \includegraphics[width=7.5cm]{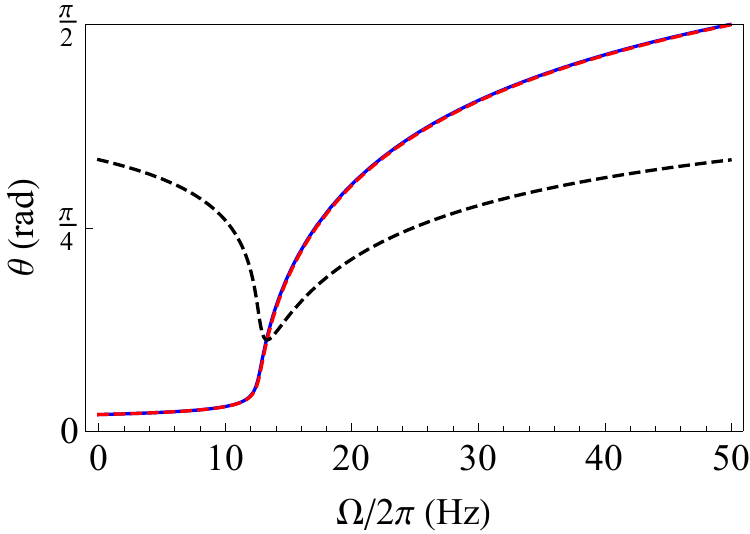}
    \caption{Upper plot: LFG precession frequency $\omega$ as a function of Larmor frequency $\Omega$; lower plot: LFG tilt angle $\theta$ as a function of $\Omega$. The Einstein-de Haas frequency is chosen to be $\omega_I = 2\pi \times 50$\,Hz and librational/nutational motion is neglected. The solid blue curves show the LFG behavior for $m\Omega_Q = 2\pi \times 49.9$\,Hz, $0.1$\,Hz below $\omega_I/\prn{2\pi}$, and the dashed red curves show the LFG behavior for $m\Omega_Q = 2\pi \times 50.1$\,Hz, $0.1$\,Hz above $\omega_I/\prn{2\pi}$. The dashed black curves illustrate the ``branch point'' condition described by Eq.\,\eqref{eq:precession-frequency-high-field-asymptote}. The equations governing the blue and red curves are obtained by simultaneously solving Eqs.\,\eqref{eq:p-phi} and \eqref{eq:quadratic-eq-for-precession-freq}.}
    \label{fig:BranchPoint}
\end{figure}

\begin{figure}
    \includegraphics[width=7.5cm]{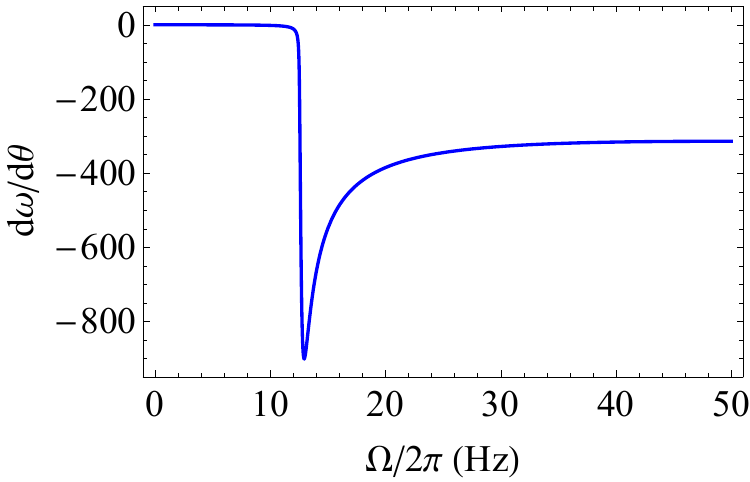}
    \caption{The derivative of the precession frequency with respect to the tilt angle, $d\omega/d\theta$, as a function of $\Omega$. The Einstein-de Haas frequency is chosen to be $\omega_I = 2\pi \times 50$\,Hz and the frequency associated with $J_z$ is chosen to be $m\Omega_Q = 2\pi \times 49.99$\,Hz.}
    \label{fig:omegaDerivative}
\end{figure}

Another example of a magnetic resonance for LFGs is encountered in the introduction to Sec.\,\ref{sec:precession-small-libration} and seen in Fig.\,\ref{fig:precession-vs-Larmor-freq}.
Key features of this phenomenon are illustrated in Fig.~\ref{fig:BranchPoint}.
The upper plot of Fig.~\ref{fig:BranchPoint} shows the LFG precession frequency $\omega$ as a function of Larmor frequency $\Omega$ for a value of $m\Omega_Q$ slightly below the Einstein-de Haas frequency $\omega_I$ (solid blue curve) and slightly above $\omega_I$ (dashed red curve), as well as a plot of the ``branch point'' condition described by Eq.\,\eqref{eq:precession-frequency-high-field-asymptote}, $\omega = \omega_I/\prn{ 2\cos\theta }$ (black dashed line).
The lower plot of Fig.~\ref{fig:BranchPoint} shows the tilt angle $\theta$ as a function of $\Omega$ for $m\Omega_Q$ slightly below and above $\omega_I$ (solid blue curve and red dashed curve, respectively), as well as the branch point condition of Eq.\,\eqref{eq:precession-frequency-high-field-asymptote}, $\theta = \cos^{-1}\sbrk{\omega_I/\prn{ 2\omega }}$ (black dashed line).
For a given projection of the total angular momentum along the magnetic field direction $J_z$, if the magnetic field $B$ is increased in the $-\hat{\bs{z}}$ direction, eventually the LFG reaches the branch point identified in Eq.\,\eqref{eq:precession-frequency-high-field-asymptote}: $\omega = \omega_I/(2\cos\theta)$.
At the branch point the relationship between the LFG precession frequency and the librational motion diverges in the no-nutation model, leading to an observable resonance in the LFG dynamics.
This can be understood by starting from Eq.\,\eqref{eq:quadratic-eq-for-precession-freq} and differentiating with respect to $\theta$, yielding
\begin{align}
-\sin\theta \omega^2 + \prn{ 2\omega \cos\theta -\omega_I } \dbyd{\omega}{\theta} = 0\,.
\end{align}
Solving for $d\omega/d\theta$, we find
\begin{align}
\dbyd{\omega}{\theta} = \frac{\omega^2\sin\theta}{2\omega \cos\theta -\omega_I }\,.
\label{eq:derivative-prec-freq-wrt-angle}
\end{align}
From Eq.\,\eqref{eq:derivative-prec-freq-wrt-angle}, we see that as $\omega \rightarrow \omega_I/(2\cos\theta)$, $\left|d\omega/d\theta\right| \rightarrow \infty$.
Thus infinitesimal changes in the tilt angle $\theta$ lead to large changes of $\omega$, implying strong coupling between libration and precession.
Within the ideal no-nutation model, the sensitivity of the precession frequency to libration diverges, but in practical LFG realizations, there will inevitably be some form of relaxation of LFG dynamics involving environmental couplings that will keep $d\omega/d\theta$ finite.
This resonance is illustrated in Fig.\,\ref{fig:omegaDerivative}.
The observed response will depend on the sweep rate through the branch point: slow sweeps in the presence of damping may allow relaxation onto a stable branch, while faster sweeps are expected to excite librational motion.

\begin{figure*}
    \centering
    \subfloat[]{\includegraphics[height=4.5cm]{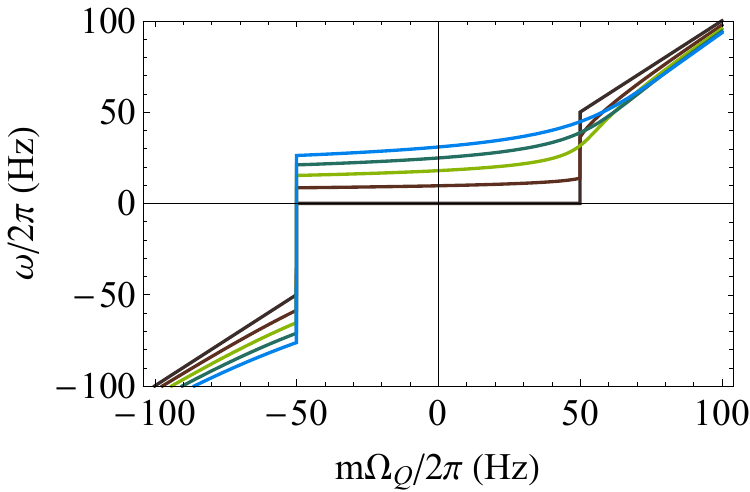}}
    \subfloat[]{\includegraphics[height=4.5cm]{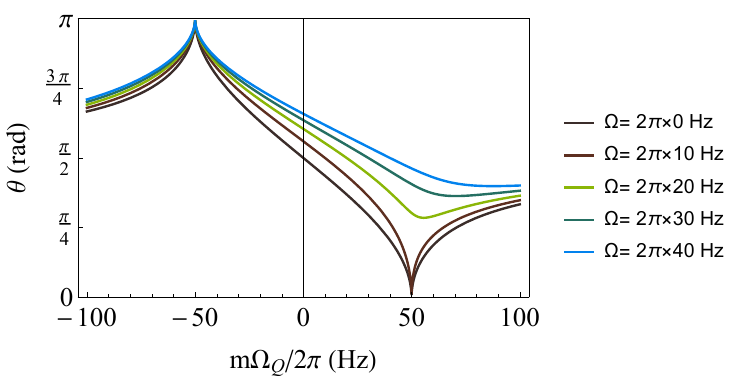}}

    \caption{Dependence of $\omega$ and $\theta$ on $m\Omega_Q$ for various values of $\Omega$. The Einstein-de Haas frequency is chosen to be $\omega_I = 2\pi \times 50$\,Hz.}
    \label{fig:DependenceOnLarmorForDiffJz}
\end{figure*}

Another special condition in the LFG dynamics accessed via rf driving of $m \rightarrow m \pm 1$ transitions is observed when near the poles.
Figure\,\ref{fig:DependenceOnLarmorForDiffJz} shows the dependence of $\omega$ and $\theta$ on $m\Omega_Q$ for various values of $\Omega < \omega_I$.
The LFG is driven to the south pole ($\theta = \pi$) as $m\Omega_Q \rightarrow -\omega_I$.
When the pole is reached, a dramatic change in the precession frequency occurs as there is a cross-over between the ``fast'' precession and ``slow'' precession regimes discussed in Sec.\,\ref{sec:precession-small-libration}.
Similar behavior is observed as the LFG is driven to the north pole ($\theta = 0$) as $m\Omega_Q \rightarrow +\omega_I$, except that if $\Omega > \omega_I/4$, the pole is avoided.
This pole avoidance can be understood as another manifestation of the branch point discussed above.
Near the north pole, define $F \equiv J_z-S$.
Expanding the effective potential for $\theta \ll 1$, one finds
\begin{align}
U\ts{eff}\prn{\theta} \approx \frac{F^2}{2I\theta^2} + \frac{S}{2} \prn{\frac{\omega_I}{4}-\Omega} \theta^2 + \cdots\,,
\end{align}
For $F \neq 0$, the first term is a centrifugal barrier that prevents the LFG from reaching the pole.
Even when $F=0$, so that the centrifugal barrier vanishes, the curvature of the effective potential at the north pole changes sign at $\Omega=\omega_I/4$.
Equivalently, evaluating the no-nutation condition, Eq.\eqref{eq:quadratic-eq-for-precession-freq}, at $\theta=0$ gives a real precession solution only for $\Omega \leq \omega_I/4$.
Thus for $\Omega>\omega_I/4$, the north-pole branch is avoided: as $m\Omega_Q$ is increased, the LFG approaches the pole only up to the branch point, then turns around in $\theta$ and continues on the other precession branch.

In addition to the branch-point resonance described above, the existence of both $m$- and $n$-ladders, described in Sec.\,\ref{subsec:ladder-spectroscopy} and shown in Fig.\,\ref{fig:equator-energy-levels}, suggests the possibility of sideband-like resonances between precessional and librational degrees of freedom.
In the absence of an rf drive, different $m$-sectors are uncoupled because $J_z$ is conserved, so crossings between $(m,n)$ levels are true crossings rather than avoided crossings.
However, an rf perturbation that contains both azimuthal and polar-angle dependence can couple states with $\Delta m = \pm 1$ and $\Delta n = \pm 1$, producing resonances whenever
\begin{align}
\omega\ts{rf} = \frac{E_{m',n'} - E_{m,n}}{\hbar}\,.
\end{align}
In the high-field equatorial regime, where $\Delta E_m \approx \hbar\Omega$ and $\Delta E_n \approx \hbar m\Omega_Q \approx \hbar\Omega$, the precessional and librational splittings become nearly equal.
This suggests that rf fields could be used not only to change $J_z$, but also to drive sideband transitions that exchange energy between precession and libration.
A detailed treatment of such avoided crossings and hybridized driven dynamics is left for future work.

We anticipate that tools for manipulation and control of LFG dynamics with rf and dc fields will open a host of possibilities for implementing advanced measurement protocols akin to those used in nuclear and electron magnetic resonance experiments \cite{claridge2016high,keeler2010understanding}.
For example, in future work we plan to investigate ensembles of LFGs where there will be inhomogeneous broadening due to, for example, varying $N$ and $I$ of the different LFGs.
In such an LFG ensemble, spin echo \cite{hahn1950spin} and related pulse sequence protocols could be useful for optimizing sensing.

%

\section{Numerical estimates}
\label{sec:numerical-estimates}

Having developed a model of LFG quantum dynamics in Secs.~\ref{sec:quantum-model} and \ref{sec:precession-small-libration}, we now turn our attention to whether such quantized dynamics are practically observable in experiments.
Table~\ref{table:numerical-parameters} summarizes numerical estimates for representative parameters for cylindrical (needle-like) LFGs of different sizes, assuming a fixed length-to-diameter aspect ratio of $\ell/d = 10$.
The choice of this aspect ratio, and the minimum length scale of $\ell = 10$\,nm in Table~\ref{table:numerical-parameters}, is made to satisfy the assumption throughout our considerations that the LFG is a single-domain ferromagnet with a well-defined macrospin $\bs{S}$ locked to the body axis \cite{kimball2016precessing}.

\begin{table*}
\caption{Estimated numerical parameters for different LFG sizes. In each case we assume a length-to-diameter aspect ratio of 10 to take advantage of the shape anisotropy to stabilize the magnetization along the long axis of the magnetic needle. We assume the needle is made of fully magnetized iron, which has an average of 2.22 polarized electron spins per atom (cobalt, for example, has similar properties, with an average of 1.72 polarized electrons per atom) \cite{chikazumi1997physics}.}
\medskip \begin{tabular}{l|c|c|c|c} \hline \hline
Parameter & \multicolumn{4}{c}{Value}  \\
\hline
\rule{0ex}{3.6ex} Length & 10\,nm & 100\,nm & 1\,$\mu$m & 10\,$\mu$m  \\
\rule{0ex}{3.6ex} Diameter & 1\,nm & 10\,nm & 100\,nm & 1\,$\mu$m  \\
\rule{0ex}{3.6ex} Mass & $6 \times 10^{-20}$\,g & $6 \times 10^{-17}$\,g & $6 \times 10^{-14}$\,g & $6 \times 10^{-11}$\,g  \\
\rule{0ex}{3.6ex} Polarized electron spins $N$ & $1.5 \times 10^3$ & $1.5 \times 10^6$ & $1.5 \times 10^9$ & $1.5 \times 10^{12}$  \\
\rule{0ex}{3.6ex} Moment of inertia $I$ & ~$5 \times 10^{-33}\,\rm{g \cdot cm^2}$~ & ~$5 \times 10^{-28}\,\rm{g \cdot cm^2}$~ & ~$5 \times 10^{-23}\,\rm{g \cdot cm^2}$~ & ~$5 \times 10^{-18}\,\rm{g \cdot cm^2}$~ \\
\rule{0ex}{3.6ex} Einstein-de Haas frequency $\omega_I/\prn{2\pi}$ & $2 \times 10^{7}$\,Hz & $2 \times 10^{5}$\,Hz & $2000$\,Hz & $20$\,Hz  \\
\rule{0ex}{3.6ex} Quantized precession frequency $\Omega_Q/\prn{2\pi}$~~~ & $3 \times 10^{4}$\,Hz & $0.3$\,Hz & $3 \times 10^{-6}$\,Hz & $3 \times 10^{-11}$\,Hz  \\
\rule{0ex}{3.6ex} Quantized magnetic field steps $\Delta B_Q$ & $0.01$\,G & $10^{-7}$\,G & $10^{-12}$\,G & $10^{-17}$\,G  \\
 ~ & ~ & ~ & ~ \\
\hline \hline
\end{tabular}
\label{table:numerical-parameters}
\end{table*}

The issue is that for sufficiently small particles at finite temperature, thermally activated magnetization reversal can occur, a phenomenon known as \emph{superparamagnetism} \cite{bean1959superparamagnetism,knobel2008superparamagnetism}.
A standard description of superparamagnetism is given by the N\'eel-Brown activation law \cite{neel1949theorie,brown1963thermal,wernsdorfer1997experimental}, where the characteristic ``reversal time'' is given by
\begin{equation}
\tau_N = \tau_0 e^{E_A/\prn{k_B T}}\,,
\label{eq:NeelBrown}
\end{equation}
where $E_A$ is the anisotropy barrier and $\tau_0$ is an ``attempt'' time (typically $\sim 10^{-10} - 10^{-9}$\,s).\footnote{The characteristic ``attempt'' time $\tau_0$ in the N\'eel-Brown model of superparamagnetism is directly related and often considered equivalent to the inverse of the characteristic frequency associated with Landau-Lifshitz-Gilbert (LLG) damping and the ferromagnetic resonance (FMR) frequency for a single-domain particle \cite{coey2010magnetism}.}
The magnetic anisotropy energy provides the potential barrier that creates the so-called \emph{magnon gap} in the magnon energy spectrum between the lowest magnon mode ($k=0$, where $k$ is the magnon wavevector) and higher magnon modes.
A particle is said to be ``blocked'' from changing its magnetization on a measurement timescale $\tau$ when $\tau_N \gg \tau$, equivalently when
\begin{equation}
E_A \gtrsim k_B T \ln(\tau/\tau_0)\,.
\label{eq:blocking_condition}
\end{equation}

For a high-aspect-ratio needle, the dominant contribution to $E_A$ is often the shape anisotropy.
In a simple demagnetizing-factor picture, the shape-anisotropy energy density for a long needle (approximated as a prolate ellipsoid) is \cite{coey2010magnetism}
\begin{equation}
K\ts{shape} = \frac{M_s^2}{4}\prn{1-3\mathcal{N}_e}\,,
\label{eq:Kshape}
\end{equation}
where $M_s$ is the saturation magnetization and $\mathcal{N}_e$ is the demagnetizing factor for the easy axis; $\mathcal{N}_e\approx 0$ for a long needle, a good approximation for an axially magnetized cylinder with aspect ratio $\ell/d=10$.
Using $M_s \approx 2\times 10^3$\,G for iron, this gives $K\ts{shape} \sim 10^6~\mathrm{erg/cm^3}$.
The anisotropy barrier is then
\begin{equation}
E_A \approx K\ts{shape} \pi r^2 \ell\ \approx K\ts{shape} \frac{\pi \ell^3}{ 400 }\,.
\label{eq:EA}
\end{equation}
For the $\ell=10$~nm LFG, $\pi r^2 \ell \approx 8 \times 10^{-21}~\rm{cm^3}$, hence $ E_A \sim 10^{-14}~\rm{erg} \sim 6~\rm{meV}$.
The corresponding reversal time for cryogenic temperatures of $T \approx 1\,\rm{K}$ is on the order of the age of the universe; for iron and the given aspect ratio:
\begin{align}
\ln\prn{ \frac{\tau_N}{\tau_0} } \approx 60 \prn{ \frac{1\,\rm{K}}{ T } } \prn{ \frac{ \ell }{ 10\,{\rm nm} } }^3\,.
\label{eq:blocking_condition_estimate}
\end{align}
Thus, for cryogenic temperatures at the 10~nm scale and above, the LFG is expected to be deeply in the blocked regime, superparamagnetic relaxation should be negligible, and the ferromagnetic state is for all practical purposes infinitely long lived.
But the exponential relationship described by Eq.\,\eqref{eq:blocking_condition_estimate} creates a dramatic scaling behavior: at a temperature of $\approx 4$\,K, the reversal time drops to a few ms!
Conversely, Eqs.~\eqref{eq:blocking_condition} and \eqref{eq:blocking_condition_estimate} enable us to estimate the minimum length scale $\ell$ required to remain blocked for at least a time of, for example, $\tau_N \gtrsim 10^6\,\rm{s}$, at a temperature of $T \sim 1\,\rm{K}$, for which we estimate $\ell \gtrsim 8\,{\rm nm}$.
This order-of-magnitude estimate supports the conclusion that the 10~nm LFG size used for the smallest example in Table~\ref{table:numerical-parameters} can ensure a single-domain, permanently magnetized ferromagnet at cryogenic temperatures, whereas smaller LFGs will tend to introduce complications due to superparamagnetic relaxation (depending, of course, on other experimental conditions).

A key qualitative result is that the quantum discreteness of the precession dynamics and energies is governed by the scale set by the quantum precession frequency $\Omega_Q=\hbar/I$, which decreases rapidly with LFG size.
The moment of inertia along the axis of a cylinder with aspect ratio 10 is
\begin{align}
I = \frac{M\ell^2}{12} = \frac{\rho \pi r^2 \ell^3}{12} \approx \frac{\rho \pi \ell^5}{4800}\,
\end{align}
where $M$ is the LFG mass and $\rho$ is the material density.
Since $\Omega_Q \propto \ell^{-5}$ (and, likewise, $\Delta B_Q \propto \ell^{-5}$) at a fixed aspect ratio, smaller LFGs are, as intuitively expected, substantially more favorable for resolving quantized precession.
The Einstein-de Haas frequency $\omega_I = S/I$, and since the number of polarized spins $N$ scales as the volume $\propto \ell^3$, $\omega_I \propto \ell^{-2}$.

For simplicity and concreteness, in the following we specifically consider equatorial precession, in which case $m\Omega_Q \approx \Omega$.
The quantized energy levels corresponding to different values of $m$ and $n$ will have the largest spacings in the high field case where $\Omega \gg \omega_I$.
The equatorial energy levels in this high-field, high-$J_z$ regime are given by (see Sec.\,\ref{subsec:equator-precession} and Appendix\,\ref{app:libration-constant-theta-SHO})
\begin{align}
E_{m,n} \approx \hbar m\Omega_Q \prn{n+\frac{1}{2}} + \frac{I}{2}\prn{ 2m\Omega_Q\Omega - \Omega^2 } \,,
\end{align}
in which case the energy splittings of the $m$-levels are
\begin{align}
\Delta E_m \approx \hbar\Omega
\end{align}
and the energy splittings of the $n$-levels are
\begin{align}
\Delta E_n \approx \hbar m\Omega_Q\,,
\end{align}
and we see that the precessional and librational energy levels have approximately the same energy splittings.
Converting to temperature via $T=\Delta E/k_B$ gives the ground state precession temperature and libration temperatures, $T_m$ and $T_n$, respectively, the temperature below which the LFG approximately occupies a single quantum energy state.
For a magnetic field of $\approx 100$\,G, where $\Omega \approx 2\pi \times 3\times 10^8$\,Hz, we estimate that
\begin{align}
T_m \approx T_n \approx 10\,{\rm mK}\,.
\end{align}
While in most experimental setups, with temperatures $\gtrsim 1$\,K, an LFG will thermally occupy a large number of $m$ and $n$ states, the above estimate shows that quantized LFG dynamics may be experimentally accessible in cryogenic experiments.

In the equatorial ladder-spectroscopy picture (Sec.\,\ref{subsubsec:Jz-spectroscopy}), driving $m \rightarrow m \pm 1$ transitions changes the expectation value of $S_z$ by an amount of order $\hbar$, implying a magnetic-moment step of order one Bohr magneton:
\begin{equation}
\Delta \mu_z \sim g\mu_B\,.
\end{equation}
This is a relatively small signal compared to the full macrospin moment $\mu \approx Ng\mu_B/2$.
Consequently, directly observing individual $m \rightarrow m \pm 1$ steps as discontinuous jumps in $\mu_z$ is challenging and will likely require averaging over many experimental cycles.

A simple geometric estimate illustrates the difficulty.
Approximating the LFG as a magnetic dipole and a pickup loop (area $A_p$) placed on-axis a distance $z$ away, the dipole field is $\approx 2\mu_z/z^3$, so the flux change associated with a $\Delta\mu_z$ step is
\begin{equation}
\Delta\Phi \sim \frac{2\Delta\mu_z}{z^3}A_p.
\label{eq:flux_step_estimate}
\end{equation}
For a micron-scale pickup loop geometry ($A_p \sim 10^{-8}\,\mathrm{cm^2}$, $z \sim 10^{-4}\,\mathrm{cm}$) and $\Delta\mu_z \sim g\mu_B$, one finds $\Delta\Phi$ to be at the level of $\sim 10^{-9}\Phi_0$, where $\Phi_0 = \hbar c/(4\pi e) \approx 2 \times 10^{-7}\,{\rm G \cdot cm^2}$ is the flux quantum.
This suggests that repeated measurements and averaging will generally be required to observe magnetization steps directly, consistent with the qualitative conclusion that the
10\,nm case is near the boundary of visibility for step-like magnetization features.
The use of nano-SQUIDs for measurement of nano-LFG dynamics is a possibility \cite{jose2017nanosquids,wernsdorfer2009micro,schmelz20173d,weber2025advanced}, in which case according to Eq.\,\eqref{eq:flux_step_estimate}, $\Delta\Phi$ would be expected to increase inversely with respect to the length scale ($\propto A_p/z^3$), potentially boosting the quantized magnetization steps into the range of visibility.

\section{Cooling}
\label{sec:cooling}

The numerical estimates of Sec.~\ref{sec:numerical-estimates} show that in order to clearly observe quantized LFG dynamics, cooling to cryogenic temperatures is required to prepare the vibrational and precessional degrees of freedom in a low-entropy state (corresponding to thermal occupation of only a few $m$ and $n$ levels).
In this section we outline a possible three-step strategy for state preparation for the case where the LFG is oriented in the polar direction along a leading field in a cryogenic, ultrahigh-vacuum environment:

\begin{enumerate}
\item[(1)] \textbf{Cool at high field:} apply a bias field $\bs{B} = B\hat{\bs{z}}$ with $B \gtrsim 100$\,G so that the LFG librational modes have a relatively large energy gap and use sideband cooling to prepare the LFG in, or near, its librational ground state;
\item[(2)] \textbf{Decouple:} turn off the sideband cooling interaction;
\item[(3)] \textbf{Adiabatically ramp down $\bs{B}$:} decrease $B$ slowly enough to preserve the librational occupation number with high probability while avoiding rethermalization by the environment.
\end{enumerate}

Alternatively, one could devise a cooling scheme for the equatorial regime (which, in fact, offers the possibility of larger spacings between energy levels at high fields, see Sec.~\ref{sec:numerical-estimates}).
Here we choose to focus on the polar case as this maps to preparing an LFG coherent state at ``rest'' as considered for precession measurements in Sec.~\ref{subsec:equator-precession}.

As discussed in Sec.~\ref{subsec:pole-precession}, near the south pole ($\theta \approx \pi$, defining the small angle $\vartheta\equiv\theta-\pi$), the Hamiltonian reduces to that of a simple harmonic oscillator for the ``exact-pole'' branch considered in Sec.\,\ref{subsec:passing-through-pole}.
As long as the condition $\abrk{J_z} \approx -S$ is fulfilled, such that the LFG is spin-dominated, the form of the Hamiltonian given by Eq.\,\eqref{eq:Hamiltonian-near-pole-varphi} remains valid even for the $\Omega \gg \omega_I$ case.
The energy gap between the ground and first excited states is $\Delta E_\ell=\hbar\omega_\ell$, and $\omega_\ell \propto \sqrt{B}$, so applying a relatively large bias field $\bs{B} = B\hat{\bs{z}}$ increases the librational energy spacing and reduces the ground-state LFG angular spread given by Eq.\,\eqref{eq:ground-state-angular-spread-at-pole}.
The comparatively large energy gap between the $n=0$ and $n=1$ states makes ground-state (or near-ground-state) preparation more accessible than in the low-field regime.
For an LFG with $\ell = 10$\,nm as considered in the prior section (see Table~\ref{table:numerical-parameters}) and an applied field of $B \approx 100$\,G, $\omega_\ell \approx 2\pi \times 10^8$\,Hz, $\Delta E_\ell \approx 3 \times 10^{-7}$\,eV, and the corresponding temperature is $T_n \approx 4$\,mK.
We choose a bias field of $B \approx 100$\,G to stay sufficiently below the ferromagnetic resonance scale in order to preserve the validity of the rigid-macrospin model (magnetization locked to the lattice axis) we have employed throughout our considerations.
It may also be interesting in future work to consider still higher fields where the rigid-macrospin model begins to cross over to a hybrid magno-mechanical regime.
In this regime, the internal degrees of freedom (magnon modes) and external degrees of freedom (libration) can become coupled, and the magnetization dynamics can enter the hybridized regime of ``magnomechanics'' \cite{zhang2016cavity,potts2021dynamical}, which may in fact be advantageous for cooling strategies.
For example, strong intrinsic dissipation channels (Landau-Lifshitz-Gilbert damping) in the magnomechanics regime can provide a bath to remove energy from external degrees of freedom.

In order to cool the LFG, we propose to implement coupling to a cold reservoir that preferentially induces $n \rightarrow n-1$ transitions (anti-Stokes processes) while suppressing $n \rightarrow n+1$ (Stokes) heating.
There have been several recent proposals to realize such a scheme \cite{kani2022intensive,asjad2023magnon,chen2025simultaneous}.
The common idea of these approaches is to use dynamical backaction (sideband) cooling, where a mechanical degree of freedom of a magnet is coupled to a driven electromagnetic and/or magnonic mode, so that anti-Stokes scattering preferentially removes energy from the system.
Operationally, the schemes optimize the detunings and spectral characteristics of the driving field to suppress Stokes (heating) processes and thereby cool the magnet, and suggest that temperatures below the mK scale may be achievable for both internal and external degrees of freedom.

After cooling, one may turn off the engineered interaction and allow the LFG to evolve freely in a cryogenic ultrahigh-vacuum environment while the bias field is ramped down.
The key requirement is that the timescale for rethermalization (dominated by external perturbations) is long compared to the duration of the ramp and the subsequent
interrogation time.

An estimate for the rethermalization time can be obtained from the collision rate of residual gas molecules, which is expected to be the dominant source of thermalization in an optimized LFG experiment.
Following Ref.~\cite{kimball2016precessing}, the gas collision rate is
\begin{equation}
\Gamma_\mathrm{col} \approx \frac{n A \bar{v}}{4},
\label{eq:Gamma_col}
\end{equation}
where $n$ is the residual gas density, $A$ is the cross-sectional area of the LFG, and $\bar{v}$ is the mean thermal speed.
For a cylindrical LFG of length $\ell = 10$\,nm and diameter $d=\ell/10$,
\begin{equation}
A \approx \frac{\ell^2}{20} \approx 5\times 10^{-14}\,\rm{cm^2}\,.
\end{equation}
Estimating a residual gas density $n \sim 10^3\,\rm{cm^{-3}}$ (which for dilution refrigerators in the mK temperature range, can be even much lower \cite{pobell2007matter})  and $\bar{v} \approx 10^3~\mathrm{cm/s}$, one finds
\begin{equation}
\Gamma_\mathrm{col} \approx 10^{-8}\,\rm{s^{-1}}\,,
\end{equation}
corresponding to a mean time between collisions $\Gamma_\mathrm{col}^{-1}$ on the order of years.
This implies that, in a dilution refrigerator environment where cryopumping is highly effective and residual gas density is expected to be extremely small, gas collisions can plausibly be made negligible on experimental timescales.
We therefore treat collisions as an upper-bound constraint and emphasize that other technical sources of rethermalization (e.g. electromagnetic pickup,
vibrations, and imperfect decoupling from readout circuitry) may become the dominant limitations.
This shows that by cooling at high field and then slowly ramping down the field, the LFG dynamics will remain out of thermodynamic equilibrium for extended times, and the effective temperature associated with the librational occupation can, in principle, be reduced substantially below the initial mK-scale preparation temperature.

We now consider an adiabatic reduction of the bias field from an initial large value $B_i$ to a smaller value $B_f$.
Since $\omega_\ell \propto \sqrt{B}$ in the high-field limit where $\Omega \gg \omega_I$, lowering $B$ compresses the librational energy ladder.
As $B \rightarrow 0$, $\omega_\ell$ approaches the finite value $\omega_I/2$ for the exact-pole fixed-$J_z$ branch considered here, see Sec.\,\ref{subsec:passing-through-pole}.
If the system remains isolated, the libration quantum number $n$ is an adiabatic invariant, so that a state with average occupation number $\bar{n}_\ell \ll 1$ at $B_i$ remains low-occupation during the ramp down of $B$.
Equivalently, the effective temperature associated with a fixed $\bar{n}_\ell$ scales as
\begin{equation}
T_\mathrm{eff} \propto \omega_\ell \propto \sqrt{\omega_I\Omega+\frac{\omega_I^2}{4}}\,.
\label{eq:Teff_scaling}
\end{equation}

The standard adiabaticity criterion for a harmonic oscillator with time-dependent frequency is
\begin{equation}
\left|\frac{\dot{\omega}_\ell}{\omega_\ell^2}\right| = \frac{\omega_I |\dot{\Omega}|}{2 \prn{ \omega_I\Omega + \omega_I^2/4 }^{3/2} } \ll 1\,.
\label{eq:adiabatic_criterion_omega}
\end{equation}
Using $\omega_\ell=\sqrt{\omega_I g \mu_B B/\hbar}$ for the high-field limit gives $\dot{\omega}_\ell=(\omega_\ell/2B)\dot{B}$, hence
\begin{equation}
\left|\frac{\dot{B}}{B}\right|\ll 2\omega_\ell(B)\,,
\label{eq:adiabatic_criterion_B}
\end{equation}
for the $\Omega \gg \omega_I$ case.
For the low-field case, $\Omega \ll \omega_I$, the adiabatic condition requires
\begin{equation}
\left|\dot{B}\right| \ll \frac{\hbar\omega_I^2}{4g\mu_B}\,.
\label{eq:adiabatic_criterion_B-low-field}
\end{equation}
This condition is easiest to satisfy at large $B$ and becomes more stringent at the end of the ramp, where $\omega_\ell$ is smallest.


In summary, preparation of the librational degree of freedom in, or close to, its ground state appears feasible in principle, providing a realistic starting point for subsequent coherent manipulation of LFG dynamics.





\section{Conclusions and future directions}
\label{sec:conclusion}



In this work we have developed a quantum model of the dynamics of a freely floating levitated ferromagnetic gyroscope (LFG), emphasizing the interplay between intrinsic spin, mechanical angular momentum, and magnetic torque.
A central principle is the conservation and quantization of the total angular momentum along the magnetic field axis ($z$), $J_z=S_z+L_z$, which, in the small nutation (libration) limit, leads naturally to discrete ladders of precessional states ($\ket{m}$, where $\hat{J}_z\ket{m} = m\hbar\ket{m}$, with $m$ the integral or half-integral projection of total angular momentum along $z$ in units of $\hbar$) and librational states (simple harmonic oscillator energy eigenstates $\ket{n}$, where $n=0,1,2,\ldots$).
The macrospin $\bs{S}$ behaves as a collective degree of freedom locked to the crystalline anisotropy axis as originally noted in Ref.\,\cite{kimball2016precessing}.
We have shown that LFG precession is not limited to the low-field regime considered in the original LFG-magnetometer proposal \cite{kimball2016precessing}, but persists into high-field regimes where the magnitude of the rotational angular momentum associated with precession can exceed the magnitude of the macrospin.
We have also analyzed the complementary quantum limits of localized, semiclassical precession wave packets and exact $J_z$-eigenstates $\ket{m}$, thereby clarifying the relationship between classical precession signals and the underlying quantized angular momentum.

The results of this paper provide a toolbox for controlling and probing LFG dynamics.
Radio-frequency fields can drive $\Delta m=\pm 1$ and $\Delta n=\pm 1$ transitions, enabling ladder spectroscopy, coherent manipulation of the tilt angle, and sideband-like coupling between precession and libration.
Associated rf-driven magnetic resonances offer new tools for magnetometry and precision torque measurements using LFGs \cite{kimball2016precessing,fadeev2021ferromagnetic,vinante2021surpassing,fadeev2021gravity,kalia2024ultralight,ahrens2025levitated,ahrens2026observation}.
The coupled dynamics also exhibit branch-point magnetic resonances, where the no (or small) nutation branches corresponding to ``fast'' and ``slow'' precession coalesce and the precession frequency becomes highly sensitive to librational motion.
Numerical estimates indicate, as expected, that quantum discreteness is most accessible for nanoscale LFGs, while high-field preparation followed by isolation and adiabatic field ramping offers a plausible route to low-occupation state preparation.
Taken together, these results establish a framework for using LFGs not only as ultrasensitive torque and magnetic-field sensors, but also as controllable mesoscopic quantum systems.

Several future research directions naturally follow from this work.
First, it will be important to extend the present freely floating model to experimentally realistic trapping potentials.
A particularly important case is levitation above superconductors \cite{wang2019dynamics,vinante2020ultralow,gieseler2020single,ahrens2025levitated,ahrens2026observation,vinante2022levitated}, where image-dipole fields and Meissner screening modify the LFG dynamics and can suppress the gyroscopic response to applied torques.
The rf-control tools developed here may provide a way to tune $J_z$, manipulate the LFG tilt angle, and recover sensitivity in regimes where superconducting back-action otherwise reduces the effective gyromagnetic response \cite{fadeev2021ferromagnetic}.
A complementary direction is to analyze charged LFGs confined in electrodynamic traps \cite{huillery2020spin,perdriat2021spin,perdriat2024rotational}, where electric fields can be used to control the center-of-mass motion while magnetic fields and rf drives control the spin and orientational degrees of freedom.
Such a platform would require extending the present theory to include micromotion, charge mobility, trap-induced torques, and possible coupling between translational and rotational motion.

More broadly, the quantized LFG dynamics identified here open several opportunities for quantum control and metrology.
Future work will investigate whether squeezing of $J_z$, $\phi$, or librational quadratures can improve magnetic-resonance linewidths or torque sensitivity, and whether measurement back-action can be evaded using phase-sensitive readout.
Coupling LFGs to SQUIDs, NV centers, microwave cavities, or atomic magnetometers may realize hybrid quantum systems in which macroscopic spin-mechanical motion is coherently interfaced with established quantum sensors.
The framework developed in this work may also be useful for analyzing proposals to generate macroscopic GHZ-like states in levitated ferromagnets \cite{ni2026macroscopic}, in which angular double-well dynamics generate coherent superpositions of spin-rotor orientations; in this context, the quantized $J_z$ ladder, rf-driven control of precession and libration, and sideband-like couplings discussed here could provide complementary tools for state preparation, recombination, and readout.
Arrays of multiple LFGs \cite{yang2026ultralight} could introduce dipole-mediated interactions, collective modes, synchronization, and possibly entanglement, providing a new resource for quantum-enhanced sensing.
Finally, the control techniques developed here may be applied to searches for exotic spin-dependent interactions \cite{fadeev2021ferromagnetic}, ultralight dark matter \cite{kalia2024ultralight}, and spin-gravity couplings \cite{fadeev2021gravity}, where LFGs offer a unique combination of high spin density, mechanical isolation, and quantum-controllable gyroscopic dynamics.

\section*{Acknowledgments}
The author would like to thank Alexander Sushkov and Dmitry Budker for enlightening discussions.
This work has been supported by the United States Air Force Office of Scientific Research under grant number FA9550-26-1-B228.
LLM artificial intelligence programs were used to cross-check results and assist in the editing of this manuscript, but all scientific results, concepts, and claims represent original work by the human author listed.

\appendix

\section{Geometric phase and the Wess-Zumino term in the LFG Lagrangian}
\label{app:Wess-Zumino-term}

The geometric phase is a phase difference acquired by a quantum system over the course of a cycle, in our considered case this is the precession of the LFG about a closed path.
Just as a magnetic field causes a moving electron to acquire a phase (the Aharonov-Bohm phase \cite{aharonov1959significance}), changing the parameters of a system slowly in time causes the quantum state to pick up a geometric phase, in our case the precession of $\bs{S}$ around the field $\bs{B}$.
The key point is that in addition to the usual phase acquired by a quantum state arising from the time-dependent Schr\"odinger equation, there is a geometric phase as well.

This geometric phase turns out to be equivalent to the system acting as a ``magnetic monopole'' \cite{aitchison1986berry}, and the geometric phase can be calculated using the associated monopole vector potential, denoted the ``Berry connection'' $\bs{A}$.
The geometric phase $\gamma_B$ is the line integral of the Berry connection:
\begin{align}
\gamma_B = \oint \bs{A}\cdot d\bs{\ell}\,.
\end{align}
For our problem of spin precession about a magnetic field along $\hat{\bs{z}}$, one possible gauge choice is \cite{zwanziger1990berry,loss1992suppression} \footnote{Another common choice is $\bs{A} = S\prn{1-\cos\theta} \bs{\nabla}\phi$; either choice leads to the same equations of motion. Note that $\bs{A}$ is related to the solid angle subtended by the trajectory of the precessing spin on the Bloch sphere.}
\begin{align}
\bs{A} = S\cos\theta \bs{\nabla}\phi = S \cot\theta \hat{\bs{\phi}}\,.
\end{align}

Consider the macrospin state (for example the coherent state described in Sec.\,\ref{subsec:equator-precession}) whose orientation is described by the unit vector
\begin{align}
\hat{\bs{n}} = \prn{\sin\theta\cos\phi,\sin\theta\sin\phi,\cos\theta}\,.
\end{align}
The Wess-Zumino term \cite{loss1992suppression} in the Lagrangian is
\begin{align}
\mc{L}\ts{WZ} = \bs{A}\cdot\dbydt{\hat{\bs{n}}}\,,
\end{align}
since in our case
\begin{align}
\dbydt{\hat{\bs{n}}} = \dot{\theta}\hat{\bs{\theta}} + \sin\theta \dot{\phi} \hat{\bs{\phi}}\,,
\end{align}
we find that
\begin{align}
\mc{L}\ts{WZ} = \bs{A}\cdot\dbydt{\hat{\bs{n}}} = S\cos\theta\dot{\phi} \,,
\end{align}
which is Eq.\,\eqref{eq:Wess-Zumino-term} from the main text.

\section{Precession with small amplitude librational oscillations}
\label{app:precession-small-amplitude-oscillations}

In this appendix we derive formulae describing precession with small amplitude nutation (libration).
We postulate a time-dependent tilt angle
\begin{align}
\theta(t) = \theta_0 + \delta \theta(t)\,,
\end{align}
where $\theta_0$ is the mean tilt angle and the perturbation $\delta \theta(t) \ll 1$ so that we can make the small angle approximation $\sin\prn{ \delta\theta } \approx \delta\theta$ and $\cos\prn{ \delta\theta } \approx 1$.
Appendix\,\ref{app:libration-constant-theta-SHO} shows that the Hamiltonian for small angle librational motion near the equator matches that of a simple harmonic oscillator, so we postulate an oscillatory solution:
\begin{align}
\delta \theta(t) = \delta\theta_0 e^{i\omega_\vartheta t}\,,
\label{eq:delta-theta-oscillatory-ansatz}
\end{align}
where $\delta\theta_0$ is the libration amplitude and $\omega_\vartheta$ is the corresponding libration frequency.
Intuiting that, due to $J_z$ being conserved, a time-dependent tilt angle $\theta(t)$ will lead to a time-dependent precession frequency $\dot{\phi} = \omega(t)$, we further postulate that
\begin{align}
\omega(t) = \omega_0 + \delta \omega(t)\,,
\end{align}
where $\omega_0$ is the mean precession frequency (obtained in the absence of libration) and that
\begin{align}
\delta \omega(t) = \delta\omega_0 e^{i\omega_\vartheta t}\,,
\label{eq:delta-omega-oscillatory-ansatz}
\end{align}
with $\delta\omega_0$ being the amplitude of precession frequency oscillation.
To show that the $\theta$ and $\omega$ oscillations are in phase as we assume in Eqs.\,\eqref{eq:delta-theta-oscillatory-ansatz} and \eqref{eq:delta-omega-oscillatory-ansatz},
consider Eq.\,\eqref{eq:Eqn-of-motion-2} in the small angle limit, keeping only terms to first order in small parameters:
\begin{align}
\sin\theta_0 \dot{\delta\omega} + 2\omega_0 \cos\theta_0 \dot{\delta\theta} - \omega_I \dot{\delta\theta} = 0\,.
\label{eq:Eqn-of-motion-2-small-oscillations}
\end{align}
Equation\,\eqref{eq:Eqn-of-motion-2-small-oscillations} shows that $\dot{\delta\omega}$ must be in-phase with $\dot{\delta\theta}$ and match the functional time-dependence, thereby justifying our ansatzes in Eqs.\,\eqref{eq:delta-theta-oscillatory-ansatz} and \eqref{eq:delta-omega-oscillatory-ansatz}.

Furthermore, by substituting the oscillatory solutions for $\delta \theta(t)$ and $\delta \omega(t)$ into Eq.\,\eqref{eq:Eqn-of-motion-2-small-oscillations}, and making use of Eq.\,\eqref{eq:quadratic-eq-for-precession-freq}, we obtain a relation between their amplitudes:
\begin{align}
\delta\omega_0 = \frac{\omega_I - 2\omega_0\cos\theta_0}{\sin\theta_0} \delta\theta_0\,.
\label{eq:relation-between-oscillation-amplitudes-of-omega-and-theta}
\end{align}

To solve for the libration oscillation frequency, we substitute our ansatzes \eqref{eq:delta-theta-oscillatory-ansatz} and \eqref{eq:delta-omega-oscillatory-ansatz} into Eq.\,\eqref{eq:Eqn-of-motion-1}, again keeping only terms to first order in small parameters, and obtain
\begin{widetext}
\begin{align}
\omega_\vartheta^2 = \omega_I \prn{\omega_0-\Omega} \cos\theta_0 + \prn{ \omega_I - 2\omega_0\cos\theta_0 }^2 - \omega_0^2\cos\prn{2\theta_0}\,.
\label{eq:general-form-of-libration-frequency-1}
\end{align}
\end{widetext}
Since we assume small libration amplitudes, we can employ Eq.\,\eqref{eq:quadratic-eq-for-precession-freq} in Eq.\,\eqref{eq:general-form-of-libration-frequency} to further simplify
\begin{align}
\omega_\vartheta^2 = \omega_0^2\sin^2\theta_0 + \prn{ \omega_I - 2\omega_0\cos\theta_0 }^2\,.
\label{eq:general-form-of-libration-frequency}
\end{align}

While Eq.\,\eqref{eq:general-form-of-libration-frequency} gives the general form of the frequency of small oscillations of $\theta$ and $\omega$, it is useful to consider limiting cases.
Near the equator ($\theta \approx \pi/2$), $\sin\theta_0 \approx 1$ and $\cos\theta_0 \approx 0$, and we obtain
\begin{align}
\omega_\vartheta \approx \pm \sqrt{\omega_I^2 + \omega_0^2}\,,
\end{align}
which matches limiting cases considered in Sec.~\ref{subsec:equator-precession}.
The mean precession frequency $\omega_0$ is discussed in various limits in the introduction of Sec.\,\ref{sec:precession-small-libration}.

Near the south pole ($\theta \approx \pi$), $\sin\theta_0 \approx 0$ and $\cos\theta_0 \approx -1$, and we find that
\begin{align}
\omega_\vartheta \approx \pm \left| \omega_I + 2\omega_0 \right| \,,
\end{align}
agreeing with Eqs.\,\eqref{eq:librational-oscillation-freq-near-pole} and \eqref{eq:precession-freq-near-south-pole-general} and the discussion in Sec.\,\ref{subsec:general-polar-dynamics}.

Based on Eq.\,\eqref{eq:relation-between-oscillation-amplitudes-of-omega-and-theta}, in the limit where $\omega_I \gg \omega_0$,
\begin{align}
\delta\omega_0 \approx \omega_I \delta\theta_0\,,
\end{align}
which shows that there can be relatively large excursions of $\omega$ even when $\delta\theta \ll 1$.
In the low-field, slow precession regime where $\omega_0 \approx \Omega$, in order to remain within the assumption of our model where $\delta\omega \ll \omega_0$, we thus require that
\begin{align}
\delta\theta_0 \ll \frac{\Omega}{\omega_I}\,.
\end{align}

\section{Approximate Hamiltonian for the small libration amplitude limit near the equator}
\label{app:libration-constant-theta-SHO}

To derive an approximate Hamiltonian for the LFG for a fixed value of $J_z = m\hbar$ near the equator $\abrk{\theta} = \theta_m \approx \pi/2$, we begin from Eq.\,\eqref{eq:libration-Hamiltonian-constant-theta} and rewrite it in terms of the small angle $\vartheta \equiv \theta-\pi/2 \ll 1$:
\begin{align}
H_m\prn{ \vartheta,p_\theta } & \approx \frac{p_\theta^2}{2I} + \frac{1}{2I\cos^2\vartheta} \prn{ m\hbar + S\sin\vartheta }^2 - S\Omega \sin\vartheta\,, \\
& \approx \frac{p_\theta^2}{2I} + \frac{1}{2I} \prn{1+\vartheta^2} \prn{ m\hbar + S\vartheta }^2 - S\Omega\vartheta\,.
\label{eq:libration-Hamiltonian-fixed-m-small-angle}
\end{align}
Expanding the second term and keeping only terms up to second order in $\vartheta$ gives
\begin{widetext}
\begin{align}
H_m\prn{ \vartheta,p_\theta } \approx \frac{p_\theta^2}{2I} + \frac{m^2\hbar^2}{2I} + \frac{1}{2I}\sbrk{\prn{ S^2 + m^2\hbar^2 }\vartheta^2 + 2\prn{ m\hbar - I\Omega } S\vartheta }\,.
\end{align}
Completing the square for the factor in square brackets in the third term in the above equation involving $\vartheta$, we find
\begin{align}
H_m\prn{ \vartheta,p_\theta } \approx \frac{p_\theta^2}{2I} + \frac{m^2\hbar^2}{2I} + \frac{1}{2I}\prn{ S^2 + m^2\hbar^2 } \prn{ \vartheta-\vartheta_0 }^2 - \frac{1}{2I}\prn{ S^2 + m^2\hbar^2 } \vartheta_0^2 \,.
\end{align}
where
\begin{align}
\vartheta_0 = \frac{1}{1+m^2\hbar^2/S^2}\prn{\frac{I\Omega - m\hbar}{S}} = \frac{1}{1+m^2\Omega_Q^2/\omega_I^2} \prn{ \frac{\Omega - m\Omega_Q }{\omega_I} }
\label{eq:phase-shift-of-precession}
\end{align}
is a magnetic-field- and $m$-dependent equilibrium tilt angle away from $\theta = \pi/2$.\footnote{The appearance of the tilt angle $\vartheta_0$ can be seen in Fig.~\ref{fig:precession-vs-Larmor-freq} as $\theta$ depends on $\Omega$; in the case shown in Fig.~\ref{fig:precession-vs-Larmor-freq}, $m=0$ and thus $\vartheta_0=0$ when $\Omega=0$. This results from the fact that exactly at the equator where $\theta = \pi/2$, $S_z=0$ and therefore $J_z = L_z$, yielding the condition that $\Omega = m\Omega_Q$.}
Observing that
\begin{align}
\hat{p}_\theta = -i\hbar \pdbyd{}{\theta} = -i\hbar \pdbyd{}{\vartheta} = \hat{p}_{\vartheta}\,,
\end{align}
and making the substitution $\omega_I = S/I$ for the Einstein-de Haas frequency, we obtain
\begin{align}
H_m\prn{ \vartheta,p_\vartheta } \approx \frac{p_\vartheta^2}{2I} + \frac{m^2\hbar^2}{2I} + \frac{I}{2}\omega_I^2 \prn{1 + \frac{m^2\hbar^2}{S^2}} \prn{ \vartheta - \vartheta_0 }^2
- \frac{I}{2}\omega_I^2 \prn{1 + \frac{m^2\hbar^2}{S^2}} \vartheta_0^2 \,,
\label{eq:libration-Hamiltonian-SHO-approx-1}
\end{align}
which can also be expressed as
\begin{align}
H_m\prn{ \vartheta,p_\vartheta } \approx \frac{p_\vartheta^2}{2I} + \frac{I}{2} \prn{\omega_I^2 + m^2\Omega_Q^2} \prn{ \vartheta - \vartheta_0 }^2
+ \frac{I}{2}m^2\Omega_Q^2 - \frac{I}{2} \prn{ \Omega - m\Omega_Q }^2 \,.
\label{eq:libration-Hamiltonian-SHO-approx-OmegaQ-form}
\end{align}
The first two terms in Eq.\,\eqref{eq:libration-Hamiltonian-SHO-approx-OmegaQ-form} are identical to the simple harmonic oscillator Hamiltonian, and thus can be quantized leading to the following energy levels
\begin{align}
E_{m,n} \approx \hbar \sqrt{ \omega_I^2 + m^2\Omega_Q^2 } \prn{n+\frac{1}{2}} + \frac{I}{2}m^2\Omega_Q^2 - \frac{I}{2} \prn{ \Omega - m\Omega_Q }^2 \,.
\label{eq:equator-energy-levels-OmegaQ-form}
\end{align}
\end{widetext}

We can think about the last two terms in Eq.\,\eqref{eq:equator-energy-levels-OmegaQ-form} in two different ways.
Note that the condition that the equilibrium tilt of the LFG remains near the equator demands that $\vartheta_0 \approx 0$, and thus, from Eq.\,\eqref{eq:phase-shift-of-precession},
\begin{align}
\Omega \approx m\Omega_Q
\label{eq:mOmegaQequalsOmega}
\end{align}
and the last term in Eq.\,\eqref{eq:equator-energy-levels-OmegaQ-form} is approximately zero.
Nevertheless, combining the last two terms of \eqref{eq:equator-energy-levels-OmegaQ-form} yields
\begin{align}
E_{m,n} \approx \hbar \omega\ts{eq} \prn{n+\frac{1}{2}} + \frac{I}{2}\prn{ 2m\Omega_Q\Omega - \Omega^2 } \,,
\label{eq:equator-energy-levels-OmegaQ-form-2}
\end{align}
where we define, based on Eq.\,\eqref{eq:equator-energy-levels-OmegaQ-form},
\begin{align}
\omega\ts{eq} \equiv \sqrt{ \omega_I^2 + m^2\Omega_Q^2 }\,.
\label{eq:equator-frequency}
\end{align}
Employing the approximation Eq.\,\eqref{eq:mOmegaQequalsOmega} yields either
\begin{align}
E_{m,n} \approx \hbar \omega\ts{eq} \prn{n+\frac{1}{2}} + \frac{I}{2}\Omega^2
\label{eq:equator-energy-levels-OmegaQ-form-3}
\end{align}
or
\begin{align}
E_{m,n} \approx \hbar \omega\ts{eq} \prn{n+\frac{1}{2}} + \frac{I}{2}m^2\Omega_Q^2\,.
\label{eq:equator-energy-levels-OmegaQ-form-4}
\end{align}

In the case where the angular momentum of the LFG is dominated by the spin $\bs{S}$, $J_z = m\hbar \ll S$, and so $m^2\hbar^2/S^2 \ll 1$ and $\omega_I \gg m\Omega_Q$, in which case $\omega\ts{eq} \approx \omega_I$ and
\begin{align}
E_{m,n} \approx \hbar \omega_I \prn{n+\frac{1}{2}} + \frac{I}{2}\prn{ 2m\Omega_Q\Omega - \Omega^2 } \,.
\label{eq:equator-energy-levels-large-spin}
\end{align}
Another regime to consider is the case where the angular momentum of the LFG greatly exceeds $S$, in which case $J_z = m\hbar \gg S$, and so $m^2\hbar^2/S^2 \gg 1$ and $\omega_I \ll m\Omega_Q$, hence $\omega\ts{eq} \approx m\Omega_Q$.
In this regime we have
\begin{align}
E_{m,n} \approx \hbar m\Omega_Q \prn{n+\frac{1}{2}} + \frac{I}{2}\prn{ 2m\Omega_Q\Omega - \Omega^2 } \,.
\label{eq:equator-energy-levels-small-spin}
\end{align}

\section{Observables near the equator}
\label{app:higher-order-terms-precession-frequency}

We can use Eq.\,\eqref{eq:precession-op} to calculate the expectation value of the precession frequency for an eigenstate $\ket{m,n}$ of $\hat{J_z}$ and the librational simple harmonic oscillator (SHO) energy as described by the quantum numbers $m$ and $n$, respectively [see Eq.\,\eqref{eq:quantized-equator-energy-levels} and also Appendix\,\ref{app:libration-constant-theta-SHO}]:
\begin{align}
\omega_{mn} & \equiv \bra{m,n} \hat{\omega} \ket{m,n} = \abrk{ \hat{\omega} }  = \frac{\abrk{\hat{J}_z} - S\abrk{\cos\hat{\theta}} }{ I \abrk{\sin^2\hat{\theta}} } \\
& = \frac{m\hbar + S \abrk{\sin\hat{\vartheta}}}{I \abrk{\cos^2\hat{\vartheta}}} \\
& \approx \frac{m\hbar + S\abrk{\hat{\vartheta}}}{I} \prn{ 1 + \abrk{\hat{\vartheta}}^2 } \\
& \approx \prn{ m\Omega_Q + \omega_I\abrk{\hat{\vartheta}} } \prn{ 1 + \abrk{\hat{\vartheta}}^2 } \\
& \approx m\Omega_Q + \omega_I\abrk{\hat{\vartheta}} + m\Omega_Q\abrk{\hat{\vartheta}^2}\,,
\label{eq:prec-freq-up-to-3rd-order-in-vartheta}
\end{align}
where, as in Appendix\,\ref{app:libration-constant-theta-SHO}, we have expanded in terms of the small angle $\vartheta \equiv \theta-\pi/2 \ll 1$, and in Eq.\,\eqref{eq:prec-freq-up-to-3rd-order-in-vartheta} we keep only terms up to second order in $\vartheta$.

Next, we note that
\begin{align}
\hat{\vartheta} = \vartheta_0 + \delta\hat{\vartheta}\,,
\end{align}
where $\delta\hat{\vartheta}$ is the angular position operator for the librational SHO and $\vartheta_0$ is given by Eq.\,\eqref{eq:phase-shift-of-precession}.
The operator $\delta\hat{\vartheta}$ can be written in terms of SHO raising and lowering operators,
\begin{align}
\delta\hat{\vartheta} = \sqrt{ \frac{\hbar}{2 I \omega\ts{eq}} } \prn{ \hat{a}^\dagger + \hat{a} }\,.
\end{align}
We find that
\begin{align}
\abrk{ \delta\hat{\vartheta} } = 0\,
\end{align}
and
\begin{align}
\abrk{ \delta\hat{\vartheta}^2 } = \frac{\hbar}{2I\omega\ts{eq}} \prn{ 2n+1 } = \frac{\Omega_Q}{2\omega\ts{eq}}\prn{ 2n+1 }\,.
\end{align}
Using the above expressions in Eq.\,\eqref{eq:prec-freq-up-to-3rd-order-in-vartheta}, we have
\begin{align}
\omega_{mn} \approx m\Omega_Q + \omega_I\vartheta_0 + m\Omega_Q \prn{ \vartheta_0^2 + \frac{\Omega_Q}{2\omega\ts{eq}}\prn{ 2n+1 } }\,.
\label{eq:omega_mn-intermediate-step-1}
\end{align}

In the spin-dominated angular momentum case where $m\Omega_Q/\omega_I \ll 1$, expanding the angular offset term of Eq.\,\eqref{eq:phase-shift-of-precession} yields
\begin{align}
\vartheta_0 & = \frac{1}{1+m^2\Omega_Q^2/\omega_I^2} \prn{ \frac{\Omega - m\Omega_Q }{\omega_I} } \\
& \approx \prn{ \frac{\Omega - m\Omega_Q }{\omega_I} } \prn{ 1 - \frac{m^2\Omega_Q^2}{\omega_I^2} } \\
& \approx \frac{\Omega - m\Omega_Q}{\omega_I} - \frac{m^2 \Omega_Q^2\Omega}{\omega_I^3} + \frac{m^3 \Omega_Q^3}{\omega_I^3}\,.
\end{align}
A similar expansion shows
\begin{align}
\vartheta_0^2 \approx \prn{ \frac{\Omega - m\Omega_Q }{\omega_I} }^2 \prn{ 1 - \frac{2m^2\Omega_Q^2}{\omega_I^2} }\,.
\end{align}
Using the above expressions in Eq.\,\eqref{eq:omega_mn-intermediate-step-1}, after some algebra, to leading order in small parameters we find

\begin{equation}
    \begin{split}
        \omega_{mn} & \approx \Omega + \frac{m\Omega_Q^2}{2\omega_I}\prn{ 2n+1 } \\
             & + \frac{m\Omega_Q\Omega^2}{\omega_I^2} - \frac{3m^2\Omega_Q^2\Omega}{\omega_I^2} + \frac{2m^3\Omega_Q^3}{\omega_I^2}\,.
    \end{split}
    \label{eq:omega_mn}
\end{equation}


Equation\,\eqref{eq:omega_mn} describes the dependence of the observable precession frequency on the quantum numbers $m$ and $n$ and on the applied magnetic field (through the dependence on the Larmor frequency $\Omega$) in the spin-dominated precession case ($\omega_I \gg \Omega$).
Another observable is the $z$-projection of the magnetization proportional to
\begin{align}
\hat{S}_z = S\cos\hat{\theta} = -S\sin\hat{\vartheta} \approx -S\hat{\vartheta}\,,
\end{align}
with expectation value
\begin{align}
\abrk{S_z} & = \bra{m,n} \hat{S}_z \ket{m,n} \approx -S\abrk{ \hat{\vartheta} } \\
& \approx S \prn{ \frac{m\Omega_Q - \Omega}{\omega_I} + \frac{m^2\Omega_Q^2\Omega}{\omega_I^3} - \frac{m^3\Omega_Q^3}{\omega_I^3} }\,.
\label{eq:S_z_mn}
\end{align}

In the opposite limit ($J_z \gg S$), where $\omega_I \ll m\Omega_Q$,
\begin{align}
\omega_{mn} \approx m\Omega_Q + \frac{\Omega_Q}{2}\prn{ 2n+1 } + \frac{\omega_I^2}{m\Omega_Q} \prn{ \frac{\Omega}{m\Omega_Q} - 1 }\,,
\label{eq:omega_mn-big-Jz}
\end{align}
and
\begin{align}
\abrk{S_z} \approx  S \frac{\omega_I}{m\Omega_Q} \prn{ 1 - \frac{\Omega}{m\Omega_Q }}\,.
\label{eq:S_z_mn-big-Jz}
\end{align}


\bibliography{spin-gyro-bib}

@book{morin2008introduction,
  title={Introduction to classical mechanics: with problems and solutions},
  author={Morin, David},
  year={2008},
  publisher={Cambridge University Press}
}

@article{kimball2016precessing,
  title={Precessing ferromagnetic needle magnetometer},
  author={Jackson Kimball, Derek F and Sushkov, Alexander O and Budker, Dmitry},
  journal={Phys. Rev. Lett.},
  volume={116},
  number={19},
  pages={190801},
  year={2016},
  publisher={APS}
}

@book{budker2008atomic,
  title={{Atomic physics: an exploration through problems and solutions}},
  author={Budker, Dmitry and Kimball, Derek and Kimball, Derek F and DeMille, David P},
  year={2008},
  publisher={Oxford University Press, USA}
}

@article{fadeev2021ferromagnetic,
  title={Ferromagnetic gyroscopes for tests of fundamental physics},
  author={Fadeev, Pavel and Timberlake, Chris and Wang, Tao and Vinante, Andrea and Band, Yehuda B and Budker, Dmitry and Sushkov, Alexander O and Ulbricht, Hendrik and Jackson Kimball, Derek F},
  journal={Quantum Sci. Technol.},
  volume={6},
  number={2},
  pages={024006},
  year={2021},
  publisher={IOP Publishing}
}

@article{loss1992suppression,
  title={Suppression of tunneling by interference in half-integer-spin particles},
  author={Loss, Daniel and DiVincenzo, David P and Grinstein, G},
  journal={Phys. Rev. Lett.},
  volume={69},
  number={22},
  pages={3232},
  year={1992},
  publisher={APS}
}

@article{arovas1988functional,
  title={Functional integral theories of low-dimensional quantum Heisenberg models},
  author={Arovas, Daniel P and Auerbach, Assa},
  journal={Phys. Rev. B},
  volume={38},
  number={1},
  pages={316},
  year={1988},
  publisher={APS}
}

@article{mentink2019quantum,
  title={{Quantum many-body dynamics of the Einstein--de Haas effect}},
  author={Mentink, Johann H and Katsnelson, MI and Lemeshko, Mikhail},
  journal={Phys. Rev. B},
  volume={99},
  number={6},
  pages={064428},
  year={2019},
  publisher={APS}
}

@article{Ein1915spinrotation1,
  title={{Experimenteller Nachweis der Ampereschen Molekularströme}},
  author={Einstein, A and de Haas, W J},
  journal={Verh. Dtsch. Phys. Ges.},
  volume={17},
  pages={152},
  year={1915}
}

@article{Ein1915spinrotation2,
  title={{Experimental proof of the existence of Ampère's molecular currents}},
  author={Einstein, A and de Haas, W J},
  journal={Koninklijke Akademie van Wetenschappen te Amsterdam, Proceedings},
  volume={18},
  pages={696},
  year={1915}
}

@article{landau1935theory,
  title={On the theory of the dispersion of magnetic permeability in ferromagnetic bodies},
  author={Landau, L and Lifshitz, Evgeny and others},
  journal={Phys. Z. Sowjetunion},
  volume={8},
  number={153},
  pages={101},
  year={1935}
}

@article{gilbert2004phenomenological,
  title={A phenomenological theory of damping in ferromagnetic materials},
  author={Gilbert, Thomas L},
  journal={IEEE transactions on magnetics},
  volume={40},
  number={6},
  pages={3443},
  year={2004}
}

@article{kambersky2007spin,
  title={{Spin-orbital Gilbert damping in common magnetic metals}},
  author={Kambersk{\`y}, V},
  journal={Phys. Rev. B},
  volume={76},
  number={13},
  pages={134416},
  year={2007},
  publisher={APS}
}

@article{xiao2005macrospin,
  title={Macrospin models of spin transfer dynamics},
  author={Xiao, Jiang and Zangwill, A and Stiles, Mark D},
  journal={Phys. Rev. B},
  volume={72},
  number={1},
  pages={014446},
  year={2005},
  publisher={APS}
}

@article{wang2019dynamics,
  title={Dynamics of a ferromagnetic particle levitated over a superconductor},
  author={Wang, Tao and Lourette, Sean and O’Kelley, Sean R and Kayci, Metin and Band, YB and Jackson Kimball, Derek F and Sushkov, Alexander O and Budker, Dmitry},
  journal={Phys. Rev. Appl.},
  volume={11},
  number={4},
  pages={044041},
  year={2019},
  publisher={APS}
}

@article{vinante2020ultralow,
  title={Ultralow mechanical damping with Meissner-levitated ferromagnetic microparticles},
  author={Vinante, Andrea and Falferi, Paolo and Gasbarri, Giulio and Setter, A and Timberlake, Christopher and Ulbricht, Hendrik},
  journal={Phys. Rev. Appl.},
  volume={13},
  number={6},
  pages={064027},
  year={2020},
  publisher={APS}
}

@article{gieseler2020single,
  title={Single-spin magnetomechanics with levitated micromagnets},
  author={Gieseler, Jan and Kabcenell, Aaron and Rosenfeld, Emma and Schaefer, JD and Safira, Arthur and Schuetz, Martin JA and Gonzalez-Ballestero, Carlos and Rusconi, Cosimo C and Romero-Isart, Oriol and Lukin, Mikhail D},
  journal={Phys. Rev. Lett.},
  volume={124},
  number={16},
  pages={163604},
  year={2020},
  publisher={APS}
}

@article{huillery2020spin,
  title={Spin mechanics with levitating ferromagnetic particles},
  author={Huillery, P and Delord, Tom and Nicolas, L and Van Den Bossche, M and Perdriat, M and Hetet, Gabriel},
  journal={Phys. Rev. B},
  volume={101},
  number={13},
  pages={134415},
  year={2020},
  publisher={APS}
}

@article{ahrens2025levitated,
  title={Levitated ferromagnetic magnetometer with energy resolution well below $\hbar$},
  author={Ahrens, Felix and Ji, Wei and Budker, Dmitry and Timberlake, Chris and Ulbricht, Hendrik and Vinante, Andrea},
  journal={Phys. Rev. Lett.},
  volume={134},
  number={11},
  pages={110801},
  year={2025},
  publisher={APS}
}

@article{ahrens2026observation,
  title={Observation of gyroscopic coupling in a nonspinning levitated ferromagnet},
  author={Ahrens, Felix and Vinante, Andrea},
  journal={Phys. Rev. Lett.},
  volume={136},
  pages={146703},
  year={2026},
  publisher={APS}
}

@article{arecchi1972atomic,
  title={Atomic coherent states in quantum optics},
  author={Arecchi, Fortunato Tito and Courtens, Eric and Gilmore, Robert and Thomas, Harry},
  journal={Phys. Rev. A},
  volume={6},
  number={6},
  pages={2211},
  year={1972},
  publisher={APS}
}

@article{brown1963thermal,
  title={Thermal fluctuations of a single-domain particle},
  author={Brown Jr, William Fuller},
  journal={Phys. Rev.},
  volume={130},
  number={5},
  pages={1677},
  year={1963},
  publisher={APS}
}

@article{band2018dynamics,
  title={Dynamics of a magnetic needle magnetometer: Sensitivity to Landau-Lifshitz-Gilbert damping},
  author={Band, YB and Avishai, Y and Shnirman, Alexander},
  journal={Phys. Rev. Lett.},
  volume={121},
  number={16},
  pages={160801},
  year={2018},
  publisher={APS}
}

@book{chikazumi1997physics,
  title={Physics of ferromagnetism},
  author={Chikazumi, S{\=o}shin and Graham, Chad D},
  year={1997},
  publisher={{Oxford University Press}}
}

@article{frait1965ferromagnetic,
  title={{Ferromagnetic resonance in metals. Frequency dependence}},
  author={Frait, Zdenek and MacFaden, Harold},
  journal={Phys. Rev.},
  volume={139},
  number={4A},
  pages={1173},
  year={1965},
  publisher={APS}
}

@article{diehl2001crystalline,
  title={Crystalline, shape, and surface anisotropy in two crystal morphologies of superparamagnetic cobalt nanoparticles by ferromagnetic resonance},
  author={Diehl, Michael R and Yu, Jae-Young and Heath, James R and Held, Glenn A and Doyle, Hugh and Sun, Shouheng and Murray, Christopher B},
  journal={J. Phys. Chem. B},
  volume={105},
  number={33},
  pages={7913},
  year={2001},
  publisher={ACS Publications}
}

@article{seynaeve2001transition,
  title={Transition from a single-domain to a multidomain state in mesoscopic ferromagnetic Co structures},
  author={Seynaeve, Eric and Rens, Gunther and Volodin, AV and Temst, Kristiaan and Van Haesendonck, Christian and Bruynseraede, Yvan},
  journal={J. Appl. Phys.},
  volume={89},
  pages={531},
  year={2001},
  publisher={American Institute of Physics}
}

@article{ni2025microscopic,
  title={Microscopic theory of a precessing ferromagnet for ultrasensitive magnetometry},
  author={Ni, Xueqi and Zou, Zhixing and Lecamwasam, Ruvi and Vinante, Andrea and Budker, Dmitry and Lam, Ping Koy and Wang, Tao and Gong, Jiangbin},
  journal={Phys. Rev. Research},
  volume={7},
  number={4},
  pages={043120},
  year={2025},
  publisher={APS}
}

@article{zwanziger1990berry,
  title={Berry's phase},
  author={Zwanziger, Josef W and Koenig, Marianne and Pines, Alex},
  journal={Annu. Rev. Phys. Chem.},
  volume={41},
  pages={601},
  year={1990}
}

@article{berry1984quantal,
  title={Quantal phase factors accompanying adiabatic changes},
  author={Berry, Michael Victor},
  journal={Proc. R. Soc. Lond. A. Math. Phys. Sci.},
  volume={392},
  number={1802},
  pages={45},
  year={1984},
  publisher={The Royal Society London}
}

@article{aharonov1959significance,
  title={Significance of electromagnetic potentials in the quantum theory},
  author={Aharonov, Yakir and Bohm, David},
  journal={Phys. Rev.},
  volume={115},
  pages={485},
  year={1959},
  publisher={APS}
}

@article{aitchison1986berry,
  title={{Berry phases, magnetic monopoles, and Wess-Zumino terms: or how the skyrmion got its spin}},
  author={Aitchison, Ian Johnston Rhind},
  journal={Acta Phys. Pol. B},
  volume={18},
  pages={207},
  year={1986}
}

@book{clarke2006squid,
  title={The SQUID handbook: Applications of SQUIDs and SQUID systems},
  author={Clarke, John and Braginski, Alex I},
  year={2006},
  publisher={John Wiley \& Sons}
}

@article{kani2022intensive,
  title={Intensive cavity-magnomechanical cooling of a levitated macromagnet},
  author={Kani, A and Sarma, B and Twamley, J},
  journal={Phys. Rev. Lett.},
  volume={128},
  pages={013602},
  year={2022},
  publisher={APS}
}

@article{chen2025simultaneous,
  title={Simultaneous cooling of the internal and external degrees of freedom of a levitated micromagnet in a cavity magnomechanical system},
  author={Chen, Lei and Liu, Yang and Bin, Liang and Ye, Sai-Yun and Zhong, Zhi-Rong},
  journal={Phys. Rev. Research},
  volume={7},
  pages={033157},
  year={2025},
  publisher={APS}
}

@article{asjad2023magnon,
  title={Magnon squeezing enhanced ground-state cooling in cavity magnomechanics},
  author={Asjad, M and Li, Jie and Zhu, Shi-Yao and You, JQ},
  journal={Fundamental Research},
  volume={3},
  pages={3},
  year={2023},
  publisher={Elsevier}
}

@article{bean1959superparamagnetism,
  title={Superparamagnetism},
  author={Bean, C P and Livingston, J D},
  journal={J. Appl. Phys.},
  volume={30},
  pages={S120},
  year={1959},
  publisher={American Institute of Physics}
}

@article{knobel2008superparamagnetism,
  title={Superparamagnetism and other magnetic features in granular materials: a review on ideal and real systems},
  author={Knobel, Marcelo and Nunes, WC and Socolovsky, LM and De Biasi, Emilio and Vargas, JM and Denardin, JC},
  journal={J. Nanosci. Nanotechnol.},
  volume={8},
  pages={2836},
  year={2008},
  publisher={American Scientific Publishers}
}

@article{wernsdorfer1997experimental,
  title={{Experimental evidence of the N{\'e}el-Brown model of magnetization reversal}},
  author={Wernsdorfer, W and Orozco, E Bonet and Hasselbach, Klaus and Benoit, A and Barbara, B and Demoncy, N and Loiseau, A and Pascard, H and Mailly, D},
  journal={Phys. Rev. Lett.},
  volume={78},
  number={9},
  pages={1791},
  year={1997},
  publisher={APS}
}

@inproceedings{neel1949theorie,
  title={{Th{\'e}orie du tra{\^\i}nage magn{\'e}tique des ferromagn{\'e}tiques en grains fins avec application aux terres cuites}},
  author={N{\'e}el, Louis},
  booktitle={Annales de g{\'e}ophysique},
  volume={5},
  pages={99},
  year={1949}
}

@book{coey2010magnetism,
  title={Magnetism and magnetic materials},
  author={Coey, John MD},
  year={2010},
  publisher={{Cambridge University Press}}
}

@article{jose2017nanosquids,
  title={NanoSQUIDs: Basics \& recent advances},
  author={Jos{\'e} Mart{\'\i}nez-P{\'e}rez, Maria and Koelle, Dieter},
  journal={Phys. Sci. Rev.},
  volume={2},
  pages={20175001},
  year={2017},
  publisher={De Gruyter}
}

@article{wernsdorfer2009micro,
  title={From micro-to nano-SQUIDs: applications to nanomagnetism},
  author={Wernsdorfer, Wolfgang},
  journal={Superconductor Science and Technology},
  volume={22},
  pages={064013},
  year={2009}
}

@article{schmelz20173d,
  title={3D nanoSQUID based on tunnel nano-junctions with an energy sensitivity of 1.3 h at 4.2 K},
  author={Schmelz, M and Vettoliere, A and Zakosarenko, V and De Leo, N and Fretto, M and Stolz, R and Granata, C},
  journal={Appl. Phys. Lett.},
  volume={111},
  year={2017},
  publisher={AIP Publishing}
}

@article{weber2025advanced,
  title={Advanced SQUID-on-lever scanning probe for high-sensitivity magnetic microscopy with sub-100-nm spatial resolution},
  author={Weber, Timur and Jetter, Daniel and Ullmann, Jan and Koch, Simon A and Pfander, Simon F and Kress, Katharina and Vervelaki, Andriani and Gross, Boris and Kieler, Oliver and Drechsler, Ute and others},
  journal={Phys. Rev. Appl.},
  volume={24},
  pages={054041},
  year={2025},
  publisher={APS}
}

@article{zhang2016cavity,
  title={Cavity magnomechanics},
  author={Zhang, Xufeng and Zou, Chang-Ling and Jiang, Liang and Tang, Hong X},
  journal={Sci. Adv.},
  volume={2},
  pages={e1501286},
  year={2016},
  publisher={American Association for the Advancement of Science}
}

@article{potts2021dynamical,
  title={Dynamical backaction magnomechanics},
  author={Potts, Clinton A and Varga, Emil and Bittencourt, Victor ASV and Kusminskiy, S Viola and Davis, John P},
  journal={Phys. Rev. X},
  volume={11},
  pages={031053},
  year={2021},
  publisher={APS}
}

@book{pobell2007matter,
  title={Matter and methods at low temperatures},
  author={Pobell, Frank},
  volume={2},
  year={2007},
  publisher={Springer}
}

@book{keeler2010understanding,
  title={{Understanding NMR spectroscopy}},
  author={Keeler, James},
  year={2010},
  publisher={John Wiley \& Sons}
}

@book{claridge2016high,
  title={{High-resolution NMR techniques in organic chemistry}},
  author={Claridge, Timothy DW},
  volume={27},
  year={2016},
  publisher={Elsevier}
}

@article{hahn1950spin,
  title={Spin echoes},
  author={Hahn, Erwin L},
  journal={Phys. Rev.},
  volume={80},
  pages={580},
  year={1950},
  publisher={APS}
}

@article{carruthers1968phase,
  title={Phase and angle variables in quantum mechanics},
  author={Carruthers, Peter and Nieto, Michael Martin},
  journal={Rev. Mod. Phys.},
  volume={40},
  pages={411},
  year={1968},
  publisher={APS}
}

@article{judge1963uncertainty,
  title={{On the uncertainty relation for $L_z$ and $\phi$}},
  author={Judge, D},
  journal={Phys. Lett.},
  volume={5},
  year={1963},
  publisher={University Coll., Dublin; and Dublin Inst. for Advanced Studies}
}

@article{robertson1929uncertainty,
  title={The uncertainty principle},
  author={Robertson, Howard Percy},
  journal={Phys. Rev.},
  volume={34},
  pages={163},
  year={1929},
  publisher={APS}
}

@article{vinante2021surpassing,
  title={{Surpassing the Energy Resolution Limit with ferromagnetic torque sensors}},
  author={Vinante, Andrea and Timberlake, Chris and Budker, Dmitry and Jackson Kimball, Derek F and Sushkov, Alexander O and Ulbricht, Hendrik},
  journal={Phys. Rev. Lett.},
  volume={127},
  number={7},
  pages={070801},
  year={2021},
  publisher={APS}
}

@article{fadeev2021gravity,
  title={Gravity Probe Spin: Prospects for measuring general-relativistic precession of intrinsic spin using a ferromagnetic gyroscope},
  author={Fadeev, Pavel and Wang, Tao and Band, YB and Budker, Dmitry and Graham, Peter W and Sushkov, Alexander O and Jackson Kimball, Derek F},
  journal={Phys. Rev. D},
  volume={103},
  number={4},
  pages={044056},
  year={2021},
  publisher={APS}
}

@article{kalia2024ultralight,
  title={Ultralight dark matter detection with levitated ferromagnets},
  author={Kalia, Saarik and Budker, Dmitry and Kimball, Derek F Jackson and Ji, Wei and Liu, Zhen and Sushkov, Alexander O and Timberlake, Chris and Ulbricht, Hendrik and Vinante, Andrea and Wang, Tao},
  journal={Phys. Rev. D},
  volume={110},
  number={11},
  pages={115029},
  year={2024},
  publisher={APS}
}

@article{perdriat2021spin,
  title={Spin-mechanics with nitrogen-vacancy centers and trapped particles},
  author={Perdriat, Maxime and Pellet-Mary, Cl{\'e}ment and Huillery, Paul and Rondin, Lo{\"\i}c and H{\'e}tet, Gabriel},
  journal={Micromachines},
  volume={12},
  pages={651},
  year={2021},
  publisher={MDPI}
}

@article{vinante2022levitated,
  title={Levitated micromagnets in superconducting traps: A new platform for tabletop fundamental physics experiments},
  author={Vinante, Andrea and Timberlake, Chris and Ulbricht, Hendrik},
  journal={Entropy},
  volume={24},
  pages={1642},
  year={2022},
  publisher={MDPI}
}

@article{perdriat2024rotational,
  title={Rotational locking of charged microparticles in quadrupole ion traps},
  author={Perdriat, Maxime and Rusconi, Cosimo C and Delord, Tom and Huillery, Paul and Pellet-Mary, Cl{\'e}ment and Durand, Alrik and Stickler, Benjamin A and H{\'e}tet, Gabriel},
  journal={Phys. Rev. Lett.},
  volume={133},
  pages={253602},
  year={2024},
  publisher={APS}
}

@article{belovs2025gyromagnetic,
  title={Gyromagnetic effects in dynamics of magnetic microparticles},
  author={Belovs, M and Livanovics, R and C{\=e}bers, A},
  journal={J. Magn. Magn. Mater.},
  volume={614},
  pages={172735},
  year={2025},
  publisher={Elsevier}
}

@article{ni2026macroscopic,
  title={Macroscopic Spin GHZ States with a Levitated Ferromagnet},
  author={Ni, Xueqi and Zou, Zhixing and Lam, Ping Koy and Wang, Tao and Gong, Jiangbin},
  journal={arXiv:2606.03676},
  year={2026}
}

@article{jackson2023probing,
  title={Probing fundamental physics with spin-based quantum sensors},
  author={Jackson Kimball, Derek F and Budker, Dmitry and Chupp, Timothy E and Geraci, Andrew A and Kolkowitz, Shimon and Singh, Jaideep T and Sushkov, Alexander O},
  journal={Phys. Rev. A},
  volume={108},
  pages={010101},
  year={2023},
  publisher={APS}
}

@article{palacios2022single,
  title={{Single-domain Bose condensate magnetometer achieves energy resolution per bandwidth below $\hbar$}},
  author={Palacios Alvarez, Silvana and Gomez, Pau and Coop, Simon and Zamora-Zamora, Roberto and Mazzinghi, Chiara and Mitchell, Morgan W},
  journal={Proc. Natl. Acad. Sci.},
  volume={119},
  number={6},
  pages={e2115339119},
  year={2022},
  publisher={National Acad Sciences}
}

@article{mitchell2020colloquium,
  title={Colloquium: Quantum limits to the energy resolution of magnetic field sensors},
  author={Mitchell, Morgan W and Alvarez, Silvana Palacios},
  journal={Rev. Mod. Phys.},
  volume={92},
  number={2},
  pages={021001},
  year={2020},
  publisher={APS}
}

@article{wachter2026gyroscopically,
  title={Gyroscopically stabilized quantum spin rotors},
  author={Wachter, Vanessa and Kusminskiy, Silvia Viola and H{\'e}tet, Gabriel and Stickler, Benjamin A},
  journal={Phys. Rev. Lett.},
  volume={136},
  pages={073604},
  year={2026},
  publisher={APS}
}

@article{stickler2021quantum,
  title={Quantum rotations of nanoparticles},
  author={Stickler, Benjamin A and Hornberger, Klaus and Kim, MS},
  journal={Nat. Rev. Phys.},
  volume={3},
  pages={589},
  year={2021},
  publisher={Nature Publishing Group UK London}
}

@article{barry2023ferrimagnetic,
  title={Ferrimagnetic oscillator magnetometer},
  author={Barry, John F and Irion, Reed A and Steinecker, Matthew H and Freeman, Daniel K and Kedziora, Jessica J and Wilcox, Reginald G and Braje, Danielle A},
  journal={Phys. Rev. Appl.},
  volume={19},
  pages={044044},
  year={2023},
  publisher={APS}
}

@article{fuwa2023ferromagnetic,
  title={Ferromagnetic levitation and harmonic trapping of a milligram-scale yttrium iron garnet sphere},
  author={Fuwa, Maria and Sakagami, Ryosuke and Tamegai, Tsuyoshi},
  journal={Phys. Rev. A},
  volume={108},
  pages={063511},
  year={2023},
  publisher={APS}
}

@article{janse2024characterization,
  title={Characterization of a levitated sub-milligram ferromagnetic cube in a planar alternating-current magnetic Paul trap},
  author={Janse, Martijn and van der Bent, Eli and Laurman, Mart and Smit, Robert and Hensen, Bas},
  journal={Appl. Phys. Lett.},
  volume={125},
  year={2024},
  publisher={AIP Publishing}
}

@article{ji2025levitated,
  title={Levitated sensor for magnetometry in ambient environment},
  author={Ji, Wei and Xu, Changhao and Qu, Guofeng and Budker, Dmitry},
  journal={arXiv:2504.21524},
  year={2025}
}

@article{yang2026ultralight,
  title={Ultralight Dark Matter Detection with a Ferromagnet Lattice},
  author={Yang, Dongyi and Yang, Xiao and Sun, Chenxi and Zhang, Jianwei},
  journal={arXiv:2602.17291},
  year={2026}
}


\end{document}